\documentclass[11pt]{amsart}
\usepackage{fullpage}
\usepackage{setspace}
\usepackage{amsmath}
\usepackage{amssymb}
\usepackage{graphicx}
\usepackage{graphicx}
\usepackage{dcolumn}
\usepackage{bm}
\usepackage{amsfonts}
\usepackage{latexsym}
\usepackage{pdfsync}
\usepackage{epsfig}
\usepackage{subfigure}
\usepackage{textcomp}
\usepackage{color}

\begin{document}
\newcommand{\abs}[1]{\left| {}_{}^{} {#1}{}_{}^{} \right|}
\newcommand{\norm}[1]{\left\|{#1}\right\|}
\renewcommand{\vec}[1]{\boldsymbol{#1}}
\newcommand{\unitvec}[1]{\boldsymbol{\widehat{#1}}}
\newcommand{\commentout}[1]{}
\newcommand{\e}{\mathrm{e}} 
\renewcommand{\i}{{i}}
\newcommand{\nwc}{\newcommand}
\newcommand{\bz}{{\mathbf z}}
\newcommand{\sqk}{\sqrt{\ks}}
\newcommand{\sqkone}{\sqrt{|\ks_1|}}
\newcommand{\sqktwo}{\sqrt{|\ks_2|}}
\newcommand{\invsqkone}{|\ks_1|^{-1/2}}
\newcommand{\invsqktwo}{|\ks_2|^{-1/2}}
\newcommand{\partz}{\frac{\partial}{\partial z}}
\newcommand{\grady}{\nabla_{\ba}}
\newcommand{\gradp}{\nabla_{\bp}}
\newcommand{\gradx}{\nabla_{\bx}}
\newcommand{\invf}{\cF^{-1}_2}
\newcommand{\myphi}{\Phi_{(\eta,\rho)}}
\newcommand{\minrg}{|\min{(\rho,\gamma^{-1})}|}
\newcommand{\al}{\alpha}
\newcommand{\xvec}{\vec{\mathbf x}}
\newcommand{\kvec}{{\vec{\mathbf k}}}
\newcommand{\lt}{\left}
\newcommand{\ksq}{\sqrt{\ks}}
\newcommand{\rt}{\right}
\newcommand{\ga}{\gamma}
\newcommand{\vas}{\varepsilon}
\newcommand{\lan}{\left\langle}
\newcommand{\ran}{\right\rangle}
\newcommand{\tvas}{{W_z^\vas}}
\newcommand{\psiep}{{W_z^\vas}}
\newcommand{\wep}{{W^\vas}}
\newcommand{\weptil}{{\tilde{W}^\vas}}
\newcommand{\wepz}{{W_z^\vas}}
\newcommand{\weps}{{W_s^\ep}}
\newcommand{\wepsp}{{W_s^{\ep'}}}
\newcommand{\wepzp}{{W_z^{\vas'}}}
\newcommand{\wepztil}{{\tilde{W}_z^\vas}}
\newcommand{\vvas}{{\tilde{\ml L}_z^\vas}}
\newcommand{\veptil}{{\tilde{\ml L}_z^\vas}}
\newcommand{\cvc}{{{\ml L}^{\ep*}_z}}
\newcommand{\cvcp}{{{\ml L}^{\ep*'}_z}}
\newcommand{\cvp}{{{\ml L}^{\ep*'}_z}}
\newcommand{\cvtil}{{\tilde{\ml L}^{\ep*}_z}}
\newcommand{\cvtilp}{{\tilde{\ml L}^{\ep*'}_z}}
\newcommand{\vtil}{{\tilde{V}^\ep_z}}
\newcommand{\ktil}{\tilde{K}}
\newcommand{\n}{\nabla}
\newcommand{\tkappa}{\tilde\kappa}
\newcommand{\ks}{{\omega}}
\newcommand{\bx}{\mb x}
\newcommand{\br}{\mb r}
\nwc{\bR}{\mb R}
\nwc{\bH}{{\mb H}}
\newcommand{\bu}{\mathbf u}
\nwc{\bxp}{{{\mathbf x}}}
\nwc{\bap}{{{\mathbf y}}}
\newcommand{\bD}{\mathbf D}
\newcommand{\bA}{\mathbf \Phi}
\nwc{\bPhi}{\mathbf{\Phi}}
\nwc{\bPsi}{\mathbf{\Psi}}
\nwc{\bh}{\mathbf h}
\newcommand{\bB}{\mathbf B}
\newcommand{\bC}{\mathbf C}
\newcommand{\bp}{\mathbf p}
\newcommand{\bq}{\mathbf q}
\nwc{\bI}{\mathbf I}
\nwc{\bP}{\mathbf P}
\nwc{\bs}{\mathbf s}
\nwc{\bd}{\mathbf{d}}
\nwc{\bX}{\mathbf X}
\newcommand{\pdg}{\bp\cdot\nabla}
\newcommand{\pdgx}{\bp\cdot\nabla_\bx}
\newcommand{\one}{1\hspace{-4.4pt}1}
\newcommand{\corr}{r_{\eta,\rho}}
\newcommand{\rinf}{r_{\eta,\infty}}
\newcommand{\rzero}{r_{0,\rho}}
\newcommand{\rzeroinf}{r_{0,\infty}}
\nwc{\om}{\omega}

\nwc{\nwt}{\newtheorem}
\nwc{\xp}{{x^{\perp}}}
\nwc{\yp}{{y^{\perp}}}
\nwt{remark}{Remark}
\nwt{definition}{Definition} 
\nwt{corollary}{Corollary} 

\nwc{\ba}{{\mb a}}
\nwc{\bal}{\begin{align}}
\nwc{\ben}{\begin{equation*}}
\nwc{\bea}{\begin{eqnarray}}
\nwc{\beq}{\begin{eqnarray}}
\nwc{\bean}{\begin{eqnarray*}}
\nwc{\beqn}{\begin{eqnarray*}}
\nwc{\beqast}{\begin{eqnarray*}}

\nwc{\eal}{\end{align}}
\nwc{\een}{\end{equation*}}
\nwc{\eea}{\end{eqnarray}}
\nwc{\eeq}{\end{eqnarray}}
\nwc{\eean}{\end{eqnarray*}}
\nwc{\eeqn}{\end{eqnarray*}}
\nwc{\eeqast}{\end{eqnarray*}}

\nwc{\vep}{\varepsilon}
\nwc{\ep}{\epsilon}
\nwc{\ept}{\epsilon}
\nwc{\vrho}{\varrho}
\nwc{\orho}{\bar\varrho}
\nwc{\ou}{\bar u}
\nwc{\vpsi}{\varpsi}
\nwc{\lamb}{\lambda}
\nwc{\Var}{{\rm Var}}

\nwt{proposition}{Proposition}
\nwt{theorem}{Theorem}
\nwt{summary}{Summary}
\nwt{lemma}{Lemma}
\nwt{cor}{Corollary}
\nwc{\nn}{\nonumber}
\nwc{\mf}{\mathbf}
\nwc{\mb}{\mathbf}
\nwc{\ml}{\mathcal}

\nwc{\IA}{\mathbb{A}} 
\nwc{\bi}{\mathbf i}
\nwc{\bo}{\mathbf o}
\nwc{\IB}{\mathbb{B}}
\nwc{\IC}{\mathbb{C}} 
\nwc{\ID}{\mathbb{D}} 
\nwc{\IM}{\mathbb{M}} 
\nwc{\IP}{\mathbb{P}} 
\nwc{\II}{\mathbb{I}} 
\nwc{\IE}{\mathbb{E}} 
\nwc{\IF}{\mathbb{F}} 
\nwc{\IG}{\mathbb{G}} 
\nwc{\IN}{\mathbb{N}} 
\nwc{\IQ}{\mathbb{Q}} 
\nwc{\IR}{\mathbb{R}} 
\nwc{\IT}{\mathbb{T}} 
\nwc{\IZ}{\mathbb{Z}} 

\nwc{\cE}{{\ml E}}
\nwc{\cP}{{\ml P}}
\nwc{\cQ}{{\ml Q}}
\nwc{\cL}{{\ml L}}
\nwc{\cX}{{\ml X}}
\nwc{\cW}{{\ml W}}
\nwc{\cZ}{{\ml Z}}
\nwc{\cR}{{\ml R}}
\nwc{\cV}{{\ml V}}
\nwc{\cT}{{\ml T}}
\nwc{\crV}{{\ml L}_{(\delta,\rho)}}
\nwc{\cC}{{\ml C}}
\nwc{\cO}{{\ml O}}
\nwc{\cA}{{\ml A}}
\nwc{\cK}{{\ml K}}
\nwc{\cB}{{\ml B}}
\nwc{\cD}{{\ml D}}
\nwc{\cF}{{\ml F}}
\nwc{\cS}{{\ml S}}
\nwc{\cM}{{\ml M}}
\nwc{\cG}{{\ml G}}
\nwc{\cH}{{\ml H}}
\nwc{\bk}{{\mb k}}
\nwc{\bn}{{\mb n}}
\nwc{\cbz}{\overline{\cB}_z}
\nwc{\supp}{{\hbox{supp}}}
\nwc{\fR}{\mathfrak{R}}
\nwc{\bY}{\mathbf Y}
\newcommand{\mbr}{\mb r}
\nwc{\pft}{\cF^{-1}_2}
\nwc{\bU}{{\mb U}}
\nwc{\bG}{{\mb G}}
\nwc{\bg}{\mathbf{g}}
\nwc{\mbf}{\mathbf{f}}
\nwc{\mbe}{\mathbf{e}}
\nwc{\be}{\mathbf{e}}
\nwc{\Om}{\Omega}
\nwc{\ind}{\operatorname{I}}
\nwc{\mbx}{\mathbf{f}}
\nwc{\bb}{\mathbf{g}}
\nwc{\xmax}{f_{\rm max}}
\nwc{\xmin}{f_{\rm min}}
\nwc{\suppx}{\hbox{\rm supp} (\mbf)}
\nwc{\by}{\mathbf{h}}
\nwc{\bZ}{\mathbf{Z}}
\nwc{\bF}{\mathbf{F}}
\nwc{\bE}{\mathbf{E}}
\nwc{\bV}{\mathbf{V}}
\nwc{\cI}{\IZ^2_N}
\nwc{\chis}{{\chi^{\rm s}}}
\nwc{\chii}{{\chi^{\rm i}}}
\nwc{\pdfi}{{f^{\rm i}}}
\nwc{\pdfs}{{f^{\rm s}}}
\nwc{\pdfii}{{f_1^{\rm i}}}
\nwc{\pdfsi}{{f_1^{\rm s}}}
\nwc{\thetatil}{{\tilde\theta}}

\nwc{\red}{\color{red}}
\nwc{\blue}{\color{blue}}

\centerline{\bf \red Chapter 3 of Optical Compressive Imaging, Taylor \& Francis 2016}
\bigskip

\title{Compressive Sensing Theory for Optical Systems Described by a Continuous Model}
\author{Albert Fannjiang}
  \address{
   Department of Mathematics,
    University of California, Davis, CA 95616-8633}
\email{fannjiang@math.ucdavis.edu}
  \thanks{
  The research supported in part by  NSF grant DMS-1413373 and Simons Foundation grant 275037.}
  
 \commentout{ \begin{abstract}
  A brief survey on  compressive sensing applications to continuous imaging models.
\end{abstract}
}
\maketitle

\tableofcontents

\section{Introduction}
 
A monochromatic  wave $u$
propagating in a heterogeneous medium  is governed by the  Helmholtz
  equation 
  \beq
  \label{helm}
  \Delta u(\br)+\om^2(1+\nu(\br)) u(\br)=0,\quad\br\in \IR^d,\quad d=2,3
  \eeq 
  where $\nu\in \IC$ describes the medium heterogeneities. 
 For simplicity, we choose the physical units 
 such that  the wave velocity is unity and 
 the wavenumber equals the frequency $\om$.

The data used for imaging is the scattered field $u^{\rm s}=u-u^{\rm i}$
governed  by 
\beq
\label{scattered}
\Delta u^{\rm s}+\om^2 u^{\rm s}=-\om^2\nu u 
\eeq
or equivalently the Lippmann-Schwinger integral equation:
\beq
\label{exact'}
u^{\rm s}(\br)&=&\om^2\int_{\IR^3} \nu(\br') 
\lt(u^{\rm i}(\br')+u^{\rm s}(\br')\rt) G(\br, \br')d\br'. 
\eeq
Here 
\beq
\label{green}
G(\br,\br')=\left\{\begin{matrix}
{e^{i\om|\br-\br'|}\over 4\pi |\br-\br'|},&d=3\\
{i\over 4} H^{(1)}_0(\om  |\br-\br'|),&d=2
\end{matrix}\right.
\eeq
is the Green function for the background propagator $(\Delta+\om^2)^{-1}$ 
 where $H^{(1)}_0$ is the zeroth order Hankel function
 of the first kind.

We consider two far-field imaging geometries: paraxial and scattering. In the former, both the object plane and the image plane are orthogonal to the optical axis while
in the latter emission  and detection of light can take any directions.
In the former, we take  $u^{\rm s}$ as the measured data and in the latter we take
the scattering amplitudes (see \eqref{sa} below) as the measured data. 
\begin{itemize}
\item{\bf Paraxial geometry}: For simplicity, let us 
state the 2D version.
Let  $\{z=z_0\}$ be the object line and  $\{z=0\}$ the image line.   
With  $\br=( x, z_0), \br'=(x', 0)$, we have
 \beq
\label{diffract}
u^{\rm s}(x,z_0)&=&
Ce^{i\om x^2/(2z_0)} \int_{\IR} \nu(x',0) \left(u^{\rm i}(x',0)+u^{\rm s}(x',0)\right)
 e^{i\om (x')^2/(2z_0)} e^{-i\om  xx'/z_0}dx' 
\eeq
where $C$ is a complex number. \\
  \begin{figure}[t]
\begin{center}
\subfigure[Diffraction geometry]{\includegraphics[width=0.4\textwidth]{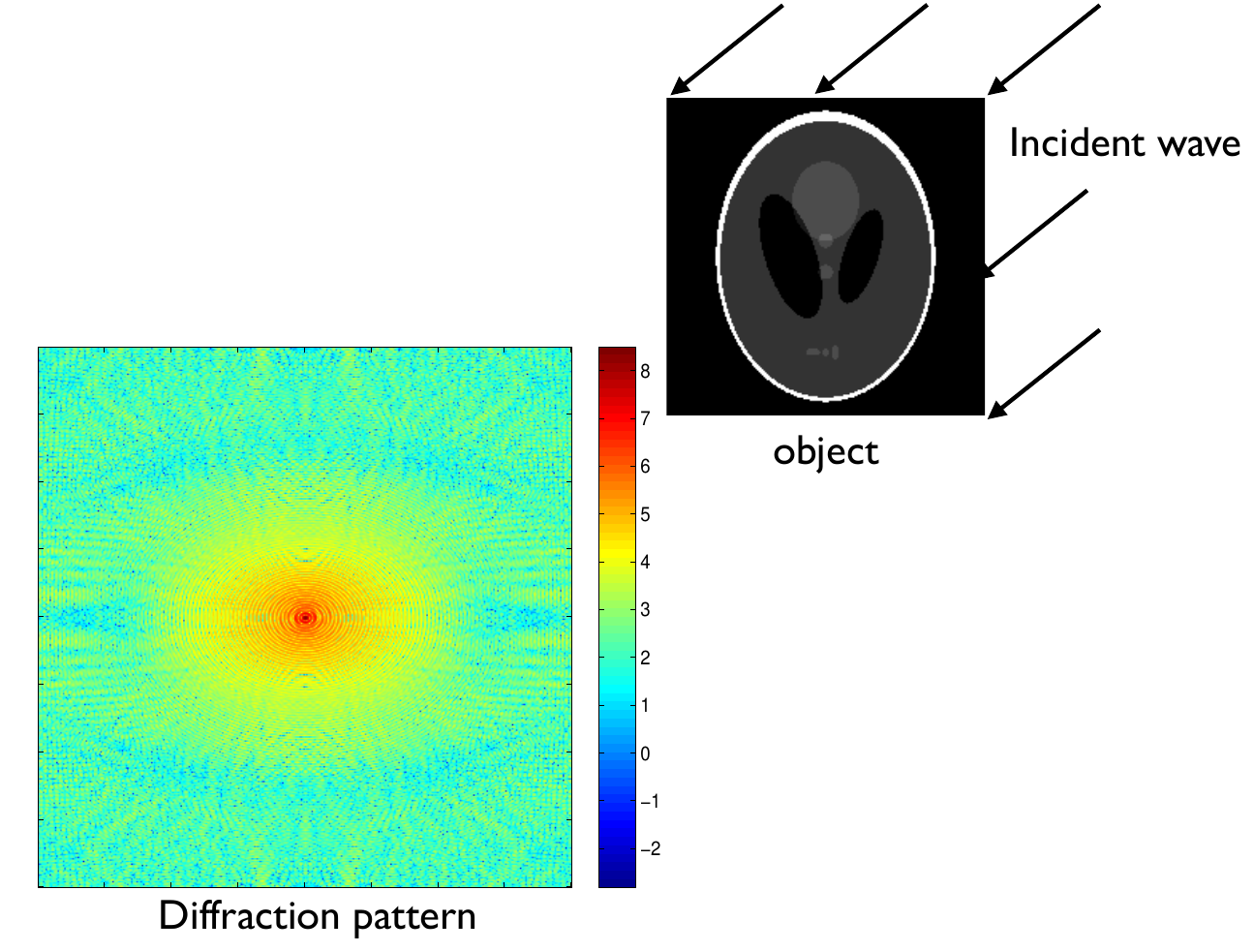}}\hspace{1cm}
\subfigure[Scattering geometry]{\includegraphics[width=8cm, height=4cm]{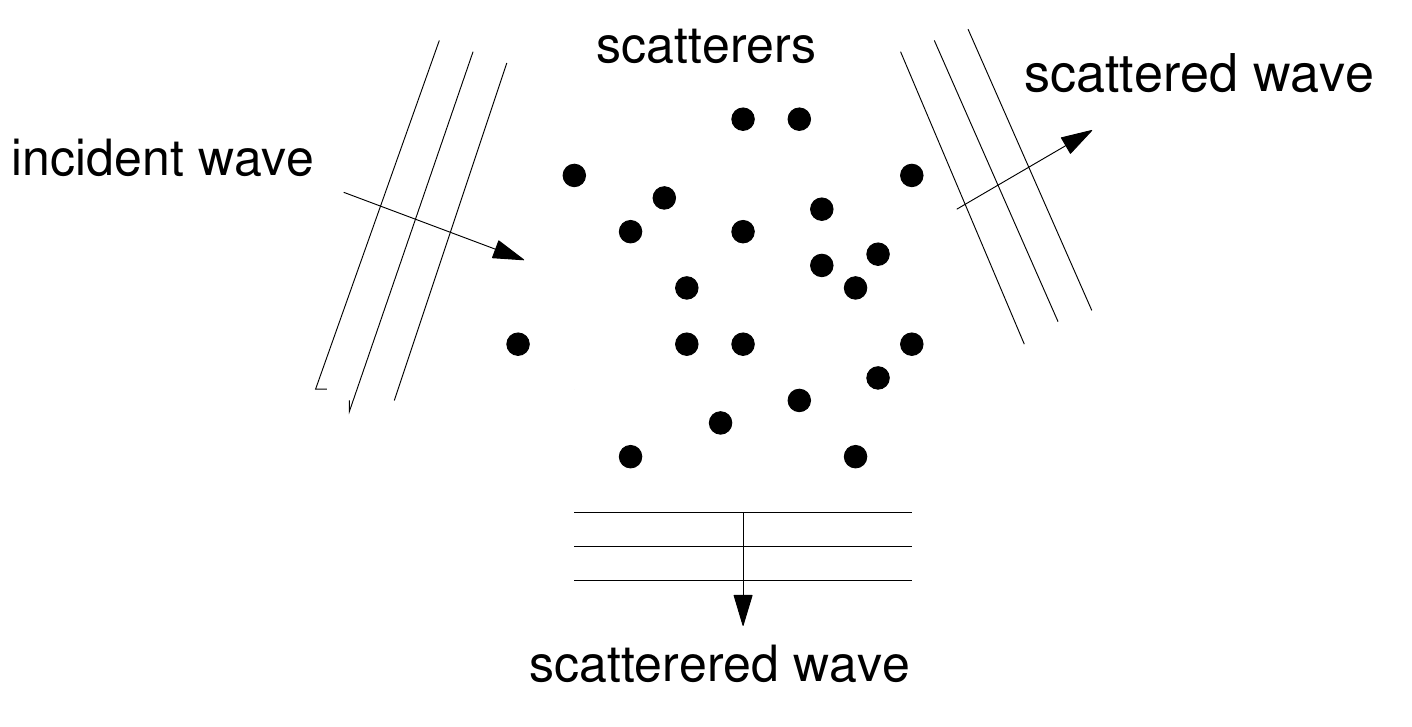}}
\end{center}
\caption{Two imaging geometries: (a) Diffraction (b) Scattering.}
\end{figure}

\item{\bf Scattering geometry:}
The scattered field has
the far-field asymptotic (Born and Wolf 1999) 
\beq
u^{\rm s}(\br)={e^{i\om |\br|}\over |\br|^{(d-1)/2}}\lt(A
(\hat\br,\hat\bd)+\cO\lt({1\over |\br| }\rt)\rt),\quad\hat\br={\br\over|\br|},\quad
d=2,3
\eeq
where the scattering amplitude $A$ has the dimension-independent form 
\beq
\label{sa}
A(\hat\br,\hat\bd)&=&{\om^2\over 4\pi}
\int_{\IR^d}  \nu(\br') \left(u^{\rm i}(\br') +u^{\rm s}(\br')\right) e^{-i\om\br'\cdot\hat\br}d\br'. 
\eeq\\
\end{itemize}
 \commentout{
With $\br=( x, z_0), \br'=(x', 0)$ and  under the paraxial approximation
 \[
\om|\br-\br'|\approx \om \lt(z_0+{|x-x'|^2\over 2z_0}\rt)
\] 
 the Green function  (\ref{green}) can be approximated by
the paraxial Green function 
\beq
\label{8-3}
{e^{i\om z_0}\over 4\pi z_0}  e^{i\om|x-x'|^2/(2z_0)}&=&{e^{i\om z_0}\over 4\pi z_0} e^{i\om x^2/(2z_0)}e^{-i\om  xx'/z_0} e^{i\om (x')^2/(2z_0)}. 
\eeq
}

Note that
since $u$ in \eqref{diffract} and (\ref{sa}) is part of the unknown due to multiple scattering, 
the inverse problem   is a nonlinear one. To deal with multiple scattering effects in compressive sensing,  it is natural to
split the inverse problem into two stages: In the  first stage we recover 
the {\em masked}  objects
\beqn
V(x)&=&  \nu(x,0) \left(u^{\rm i}(x,0)+u^{\rm s}(x,0)\right)
 e^{i\om x^2/(2z_0)},\quad\hbox{(paraxial geometry)} \\
 V(\br)&=& \nu(\br) \left(u^{\rm i}(\br) +u^{\rm s}(\br)\right), \quad\hbox{(scattering geometry)}
 \eeqn
 \commentout{the normalized quantities
 \beqn
 u^{\rm s}(x,z_0) 
 C^{-1} e^{-i\om x^2/(2z_0)},\quad
 A(\hat \br,\hat \bd)  4\pi \om^{-2}
 \eeqn
 as the respective data vectors} 
 with
 the Fourier-like integrals  in \eqref{diffract} and \eqref{sa}
 as the sensing operators. In the second stage, we recover
 the true objects from the masked objects. 
 
 For the most part of
 the article, however, we will {  focus on the first stage or}  make the Born approximation to linearize 
 the imaging problem and turn to the multiple scattering effect
only in Section \ref{sec8}. 
 
\section{Outline}

In Section \ref{sec:cs} we review the basic elements of compressive sensing theory
including basis pursuit and greedy algorithms (orthogonal matching pursuit, in particular).
We place greater emphasis on the incoherence properties than on the restricted isometry property because the former is much easier to estimate than the latter, even though the latter can also be established in several settings  as we will see throughout this article. 
One thing to keep in mind about incoherence is that it is far beyond the standard notion of
coherence parameter, which is the worst case metric (see \eqref{coherence} below). The incoherence properties are fully
expressed  in the Gram matrix of the sensing matrix, also known as  the coherence pattern. 
Second thing noteworthy  about  incoherence is that the standard performance  guarantees expressed in terms of the coherence parameter often underestimate the actual performance of algorithms. 
Its usefulness primarily lies in providing a guideline for
designing measurement schemes.

In Section \ref{sec3} we consider the Fresnel diffraction with the pixel basis.  The pixel basis, having a finite, definite size, is emphatically  {\em not} suitable for  point-like objects.  Indeed, in order to build incoherence in the sensing matrix,  it is imperative that the wavelength be shorter than the grid spacing. In other words, the pixel basis is suitable only for objects that are decomposable into ``smooth" parts relative to the wavelength. The sparsity priors then come  in two kinds: (i) there are few such parts with 1-norm as proxy (ii) there are few changes from part to part with the total variation as proxy (Section \ref{sec3.1}). In the context of Fourier measurement,
we introduce the notion of constrained joint sparsity to connect these two sparse priors
and discuss basis pursuit  (Section \ref{sec3.2}) and orthogonal matching pursuit for joint sparsity (Section \ref{sec3.3}). 

In contrast to the pixelated  objects, point objects naturally do not
live on grids. Such a problem arises in applications e.g. discrete spectral estimation among others. There is this fundamental tradeoff  in using a grid to image point objects with the standard theory of compressive sensing: the finer the grid, the better the point objects are captured but
the worse the coherence parameter becomes.  In Section \ref{sec4},  we use the notion of coherence band to analyze 
the coherence pattern and design new compressive sensing algorithms for imaging well separated,  off-grid point objects. In addition to off-grid point objects, the coherence-band techniques are also useful for
imaging objects that admit a sparse representation in highly redundant dictionaries. 
One celebrated example is the single-pixel camera discussed briefly in Section \ref{sec4.4}. 

In Section \ref{sec5}, we discuss  Fresnel diffraction with sparse representation in 
the Littlewood-Paley basis which is a slowly decaying wavelet basis in stark contrast to
the pixel basis and the point-like objects.   In this basis, the sensing matrix has a hierarchical 
structures completely decoupled over different scales. In Section \ref{sec6} we discuss
near-field diffraction in terms of angular spectrum which works out nicely with
the Fourier basis.

In Section \ref{sec7} we consider inverse scattering with the pixelated as well as point objects. Here we focus on the design of sampling schemes (Section \ref{sec7.2}) and
various coherence bounds for different schemes (Section \ref{sec7.3}). 

In Section \ref{sec8}, we discuss multiple scattering of point objects and the appropriate 
techniques for solving the  nonlinear inverse problem. The keys are the combination
of the coherence-band and the joint sparsity techniques developed earlier. 

In Section \ref{sec9}, we discuss inverse scattering with extended objects sparsely
represented in the Zernike basis. In Section \ref{sec10} we discuss interferometry 
with incoherent sources in astronomy. As a consequence of the celebrated Van Citter-Zernike theorem, the resulting sensing matrix has a similar structure to that
for  scattering with multiple inputs and outputs. The difference between them
lies in the fact that for interferometry the inputs and outputs are necessarily correlated while
for scattering the inputs and outputs can be independent. As a result,
the (in)coherence properties of interferometry are more subtle and it is an ongoing problem
to search for the optimal sensor arrays in optical interferometry in astronomy.

\section{Review of compressive sensing}\label{sec:cs}
A distinctive advantage   of compressive sensing is accounting  for the
finite, discrete nature of measurement by appropriately
discretizing the object domain.

By a slight abuse of notation, we use
$\|\cdot\|_p$ to denote the $p$-norm ($p\geq  1$) of functions as well as vectors, i.e.
\beq
\|f\|_p&=&\left(\int |f(\br)|^pd\br\right)^{1/p}, \quad f\in L^p(\IR^d)\\
\|\mbf\|_p&=&\left(\sum_{j=1}^N |f_j|^p\right)^{1/p},\quad \mbf\in \IC^{N}
\eeq
and $\|\mbf\|_0$ (the sparsity) denotes the number of nonzero components in a vector $\mbf$.

By discretizing the right hand side of \eqref{diffract} or \eqref{sa} and selecting a discrete set of data on the left hand side, we shall  rewrite the continuous models 
in the form of linear inversion 
\beq
\bg=\bPhi\mbf +\mbe\label{u1}
\eeq where the error vector 
 $\mbe\in \IC^M$ is the sum of  the external noise
 $\bn$ and the discretization error $\bd$ due to
 model mismatch. By definition, the discretization error $\bd$  is given by 
\beq
\bd=\bg-\bn-\bPhi \mbf. \label{de}
\eeq

Consider  the principle of 
basis pursuit denoising (BPDN)
\beq
\label{32}
\mbox{min}\,\, \|\bh\|_1,\quad \mbox{s.t.}\quad \|\bg-\bA \bh\|_2\leq \|\be\|_2=\ep.
\eeq
When $\ep=0$, \eqref{32} is called basis pursuit (BP).
With the right choice of the parameter $\lambda$,   BPDN is equivalent to the unconstrained 
convex program called  the Lasso (Tibshirani 1996)
 \beq\label{13}
\min_{\bz} {1\over 2} \|\bb-{  \bA} \bz\|_2^2+\lambda {  \ep}\|\bz\|_1.
\eeq
 Both
BPDN \eqref{32} and Lasso \eqref{13} are convex programs and
have numerically efficient solvers (Chen {\em et al.} 2001, Boyd and Vandenberghe 2004, Brucskstein {\em et al.} 2009).

A fundamental notion in compressed sensing under which
BP yields a unique exact solution is the restrictive isometry property (RIP) due to Cand\`es and Tao 2005. 
Precisely,  let the restricted isometry constant (RIC) $\delta_s$ be the smallest nonnegative  number such that the inequality
\[
\kappa (1-\delta_s) \|\bh\|_2^2\leq \|\bA \bh\|_2^2\leq \kappa (1+\delta_s)
\|\bh\|_2^2
\]
holds for all $\bh\in \IC^N$ of sparsity at most $ s$ and some
constant $\kappa>0$. RIP means a sufficiently small $ \delta_{2s}$ (see \eqref{ric} below). 

Now we recall a standard  performance guarantee 
under RIP. 
\begin{theorem} \label{thm:rip} (Cand\`es 2008) 
Suppose the  RIC of $\bA$
satisfies the inequality 
\beq
\label{ric}
\delta_{2s}<\sqrt{2}-1
\eeq
with $\kappa=1$. 
 Then the solution $\mbf_*$ of BPDN (\ref{32}) satisfies
 \beq
 \|\mbf_*-\mbf\|_2&\leq & C_1s^{-1/2}\|\mbf-\mbf^{(s)}\|_1+C_2\ep
 \label{15}
 \eeq
 for some constants $C_1$ and $C_2$ where
 $\mbf^{(s)}$ consists of the $s$ largest components, in magnitude, 
of $\mbf$. 
\end{theorem}
\commentout{
\begin{remark}
$\mbf^{(s)}$ is the best $s$-sparse
approximation of $\mbf$ in the sense of $L^1$-norm, i.e.
\[
\mbf^{(s)}=\hbox{argmin}\,\, \|\bh-\mbf\|_1,\quad\hbox{s.t.}
\quad \|\bh\|_0\leq s.
\]
\end{remark}
}
\begin{remark}
\label{rmk1}
For general $\kappa\neq 1$, we consider the normalized
version of (\ref{u1}) 
\[
{1\over \sqrt{\kappa}}\bg={1\over \sqrt{\kappa}}\bA \mbf+
{1\over \sqrt{\kappa}}\mbe
\]
and obtain from \eqref{15} that 
\beq
 \|\mbf_*-\mbf\|_2&\leq & C_1s^{-1/2}\|\mbf-\mbf^{(s)}\|_1+C_2{\ep\over\sqrt{\kappa}}. 
 \eeq

\end{remark}
Note however that neither BPDN or Lasso is an algorithm by itself
and there are many different algorithms for solving these convex programs. Some solvers are available on-line, e.g. YALL1
and 
the open source code {\em L1-MAGIC} ({\tt http://users.ece.gatech.edu/\~\,justin/l1magic/}).

Besides convex programs, greedy algorithms are an alternative approach to sparse recovery.
A widely known greedy algorithm is the Orthogonal Matching Pursuit (OMP)
(Davis {\em et al.} 1997, Pati {\em et al.} 1993). 
 
\begin{center}
   \begin{tabular}{l}
   \hline
   \centerline{{\bf Algorithm 1.}\quad Orthogonal Matching Pursuit (OMP)} \\ \hline
   Input: $\bA, \bb.$\\
 Initialization:  $\mbx^0 = 0, \br^0 = \bb$ and $S^0=\emptyset$ \\ 
Iteration: For  $j=1,...,s$\\
\quad {1) $i_{\rm max} = \hbox{arg}\max_{i}|\lan \br^{j-1},\Phi_i\ran | , i \notin S^{j-1} $} \\
  \quad      2) $S^{j} = S^{j-1} \cup \{i_{\rm max}\}$ \\
  \quad  3) $\mbx^j = \hbox{arg} \min_\bh \|
     \bA \bh-\bb\|_2$ s.t. \hbox{supp}($\bh$) ${ \subseteq} S^j$ \\
  \quad   4) $\br^j = \bb- \bA \mbx^j$\\
 Output: $\mbx^s$. \\
 \hline
   \end{tabular}
\end{center}

\bigskip

OMP has a performance guarantee in terms of the coherence parameter defined by
\begin{equation}
  \mu({ \bPhi}) = \max_{k\neq l} \mu(k,l),\quad
  \mu(k,l)={| \Phi_{k}^\dagger\Phi_{l} |\over \|\Phi_k\| \|\Phi_l\|}
   \label{coherence}
\end{equation}
where {  $\Phi_k$ is the $k$-th column of $\bA$}, $\mu(k,l)$ is the pairwise coherence parameter and the totality $[\mu(k,l)]$ is the {\em coherence pattern} of the sensing matrix $\bPhi$.  Here and below $\dagger$ denotes the conjugate transpose. 

\begin{theorem} \label{thm2} (Donoho {\em et al.} 2006) 
 Suppose that the sparsity $s$ of the signal vector $\mbf$ satisfies 
\begin{equation}
   \mu({  \bA})(2s-1) + 2 \frac{\|\mbe\|_2}{f_{\rm min}} < 1 
   \label{180}
\end{equation}
where $f_{\rm min} =\displaystyle \min_{k} |f_k|$. Denote by ${\mbf}_*$, the output of the OMP reconstruction. Then
\begin{itemize}
\item[(a)] $\mbf_*$ has the correct support, {  i.e. $
    \text{supp}({\mbf}_*) = \text{supp}(\mbf)
$
where $\text{supp}(\mbf)$ is the support of $\mbf$}.
\item[(b)] $\mbf_*$ approximates
the object vector in the sense that 
\beq
\label{190}
\|\mbf_*-\mbf\|_2\leq {\|\mbe\|\over \sqrt{1+\mu-\mu s}}.
\eeq

\end{itemize}
\end{theorem}
Incoherence or RIP often requires randomness in the sensing matrix
which can come from the randomness in sampling as well as in 
illumination. Between the two metrics, incoherence is far more flexible and
easier to verify for a  given sensing matrix. However, performance guarantees
in terms of the coherence parameter such as \eqref{180} of Theorem \ref{thm2} tend to be  conservative.

\section{Fresnel diffraction with pixel basis}\label{sec3}
As a first example, we consider the imaging equation  \eqref{diffract} for Fresnel diffraction.
We shall write \eqref{diffract} in the discrete form \eqref{u1} by discretizing
the right hand side of \eqref{diffract} and selecting a discrete set of
scattered field data for the left hand side. 

We approximate  the masked object   
 \beq
 \label{mo}
V(x)= 
 \nu(x)u(x,0)e^{i\om x^2/(2z_0)}
 \eeq
 by the discrete sum   on the scale  $\ell$
\beq
\label{18}
V_{\ell}(x)=\sum_{k=1}^N
b({x\over\ell}-k  ) V(\ell k),\quad V(\ell k)=\nu(\ell k)u(\ell k,0)e^{i\om \ell^2 k^2/(2z_0)}
\eeq
where  \beq
 \label{loc}
 b(x)=\left\{\begin{matrix}
 1,& x\in [-{1\over 2}, {1\over 2}]\\
 0,& \hbox{else}.
 \end{matrix}\right.
 \eeq
 is the localized pixel ``basis".  We assume that  $V_\ell$ is a good approximation of the masked object for sufficiently small $\ell$ in the sense $\lim_{\ell\to 0} \|V-V_\ell\|_1=0$.
 
  Moreover, we assume that $V_\ell$ is sparse in the sense that
 relatively few components $V(k\ell)$ are significant compared to the number of grid points $N$. Note that sparse objects in the pixel basis are {\em not} point-like.
 Point objects typically induce large gridding errors and requires techniques beyond
 standard compressive sensing reviewed in Section \ref{sec:cs} (cf. Section \ref{sec:blo}).

To proceed, we shall make the Born approximation and set $u^{\rm i}(x,0)=1$ (i.e. normal incidence of plane wave). 

Let $x_j, j=1,...,M$ be the sampling points on the image/sensor line and define
\beq
\label{xi}
\xi_j={\om \ell x_j\over 2\pi z_0},\quad j=1,...,M.
\eeq
Set the discretized, unknown vector $\mbf\in \IC^N$ as
\[
f_k =\nu(\ell k) e^{i\om \ell^2k^2/(2z_0)},\quad k=1,...,N
 \]
 and the data vector $\bg\in \IC^M$ as
 \[
 g_j= {u^{\rm s}(x_j,z_0)\over C{  \ell}\hat b(\xi_j)} e^{-i\om x_j^2/(2z_0)},\quad j=1,...,M
 \]
 where
 \beq
 \label{23'}
 \hat b(\xi)=\int b(x) e^{-i2\pi x\xi} dx={\sin{(\pi \xi)}\over \pi \xi}. 
 \eeq
As a result, \eqref{diffract} can be expressed as  \eqref{u1} with  the sensing matrix 
\beq
{  \bPhi}&=&\begin{bmatrix}
     \Phi_{1} & \ldots & \Phi_{N}
  \end{bmatrix}\in \IC^{M\times N},\quad 
\Phi_k= \lt[e^{-2\pi i\xi_j k}\rt]_{j=1}^M,\quad k=1,...,N. \label{20'} 
\eeq
A sensing matrix whose columns
have the same $2$-norm (as in \eqref{20'}) tends to enjoy better performance in 
compressive  sensing reconstruction.

When $\xi_j$ are independent uniform random variables
on $[-1/2,1/2]$, \eqref{20'} is the celebrated random partial
Fourier matrix which is among a few examples with a relatively
sharp bound on the RIP given below. 
\begin{theorem} (Rauhut 2008) 
Suppose \beq
{M\over \ln{M}}\geq c \delta^{-2}k\ln^2{k} \ln{N} \ln{1\over \ep},\quad
\ep\in (0,1)\label{250}
\eeq
for given sparsity $k$ where $c$ is an absolute constant. 
Then the restricted isometry constant 
of  the matrix (\ref{20'}) satisfies the bound
\[
\delta_k<\delta
\]
with
probability at least $1-\ep$.\label{thm3}
\end{theorem}
\begin{remark}
To apply Theorem \ref{thm3} in the context of Theorem \ref{thm:rip}
we can set  $k=2s$ and $\delta=\sqrt{2}-1$. Ineq. \eqref{250} then implies
that it would take roughly $ \cO(s)$, modulo some logarithmic factors,  amount of measurement data 
for BPDN to succeed in the sense of \eqref{15}. 

On the other hand, the coherence parameter $\mu$ typically scales as
$\cO(M^{-1/2})$ as we will see in Theorem \ref{thm4}, so, in view of 
the condition \eqref{180} in Theorem \ref{thm2}, the amount of needed data is $\cO(s^2)$, significantly larger than $\cO(s)$
for $  1\ll s\ll N$. 

While this observation is usually valid in the case of OMP,  it {  needs} not apply to other greedy algorithms such as Subspace Pursuit (BP) whose performance guarantee requires 
$\cO(s)$, up to logarithmic factor, amount of data (Dai and Milenkovic 2009). 

\end{remark}
The fact that $\xi_j$ are independent uniform random variables
on $[-1/2,1/2]$ implies  that $x_j$ are independent uniform random variables on $[-A/2, A/2]$ with 
\beq
\label{27'}
A={2\pi z_0\over \om \ell}
\eeq
in view of \eqref{xi}.  
Viewing  $\ell$ as the resolution {  length} of the imaging set-up we obtain
the resolution criterion
\beq
\label{pix}
\ell={2\pi z_0\over A\om}
\eeq
which is equivalent to the classical Abbe or Rayleigh criterion.

Now let us estimate  the discretization error vector $\bd$ in \eqref{de}.
Define  the transformation
$\cT$ by 
\[
 (\cT V)_j ={{  1}\over {  \ell }\hat b(\xi_j)} \int V(x') e^{-2\pi i \xi_j x'/\ell} dx', 
 \]
cf. (\ref{sa}).  By definition
  \[
 \bd= \cT V -\cT V_{\ell}
 \]
  we have
 \beq
 \|\bd\|_\infty\leq {\|V-V_{\ell}\|_1\over {  \ell} \min_{j}|\hat b(\xi_j)|}, \quad   \hat b(\xi)={\sin{(\pi \xi)}\over \pi \xi}. 
\label{24'}
 \eeq
 For $\xi\in [-1/2,1/2]$, $\min|\hat b(\xi)|={2/\pi}$ {  and $\max|\hat b(\xi)|=1$. Hence}
 \beq
\|\bd\|_2\leq \|\bd\|_\infty \sqrt{M}\leq {\pi\sqrt{M}\over 2{  \ell}}  \|V-V_{\ell}\|_1\label{25'}
 \eeq
{  and
  \[
{\|\bd\|_2\over \|\bg\|_2}\leq  {\pi C\sqrt{M}\|V-V_{\ell}\|_1\over 2 \sqrt{\sum_{j=1}^M |u^{\rm s}(x_j)|^2}}
 \]
 }
 which can be made arbitrarily small by setting $\ell$ sufficiently small while holding $M$ fixed and maintaining the relation \eqref{pix}.

\subsection{Total variation minimization}\label{sec3.1}
\label{sec:cjs}
If the masked object $V$ is better approximated by a piecewise (beyond the scale $\ell$)  constant function $V_\ell$, then the
sparsity prior  can be enforced by the  discrete total variation
\beqn
\|\bh\|_{\rm tv}&\equiv &\sum_j |\Delta h (j)|,\quad \Delta h(j)= {h_{j+1}-h_j} \label{dg}.
\eeqn
Instead of \eqref{32} we consider a different convex program, called total variation minimization (TV-min)
\beq
\label{tv1}
\mbox{min}\,\, \|\bh\|_{\rm tv},\quad \mbox{s.t.}\quad \|\bg-\bA \bh\|_2\leq \epsilon.
\eeq
 cf. (Cand\`es {\em et al.} 2006, Rudin and Osher 1994, Rudin {\em et al.} 1992, Chambolle 2004, Chambolle and Lions 1997). 

For two-dimensional objects $h(i,j), i,j=1,...,n$, let $\bh=(h_p)$ be the vectorized version with index $p=j+(i-1)n$. The 2D discrete (isotropic)  total variation  is given by  
\beqn
&\|\bh\|_{\rm tv}\equiv \sum_{i,j} \sqrt{|\Delta_1 h(i,j)|^2+ |\Delta_2 h(i,j)|^2},&\\
&\Delta_1 {h}(i,j)=(h(i+1,j)-h(i,j),\quad \Delta_2 {h}(i,j)= h(i,j+1)-h(i,j))&.\label{dg2}
\eeqn
\commentout{
as well as the {\em anisotropic} version 
\[
\|\bh\|_{\rm atv}\equiv \sum_{i,j} |\Delta_1 h(i,j)|+ |\Delta_2 h(i,j)|.
 \]
\cite{NW}. 
}

 \begin{figure}[t]
\begin{center}
\includegraphics[width=0.5\textwidth]{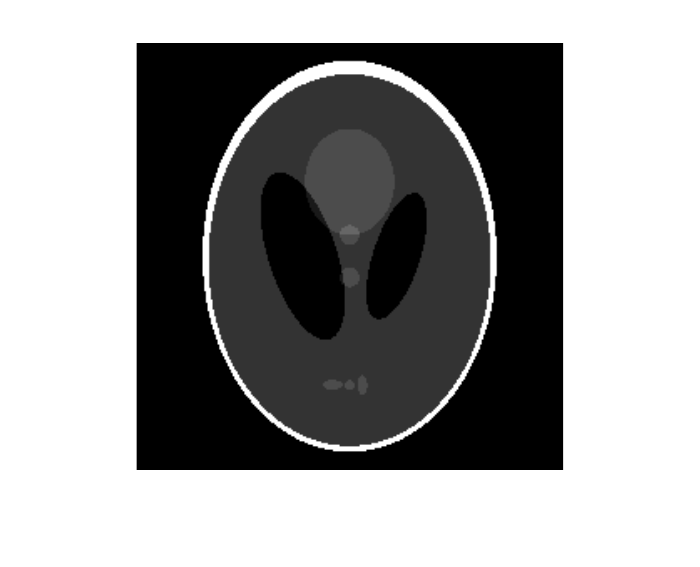}\hspace{-1.5cm}
\includegraphics[width=0.5\textwidth]{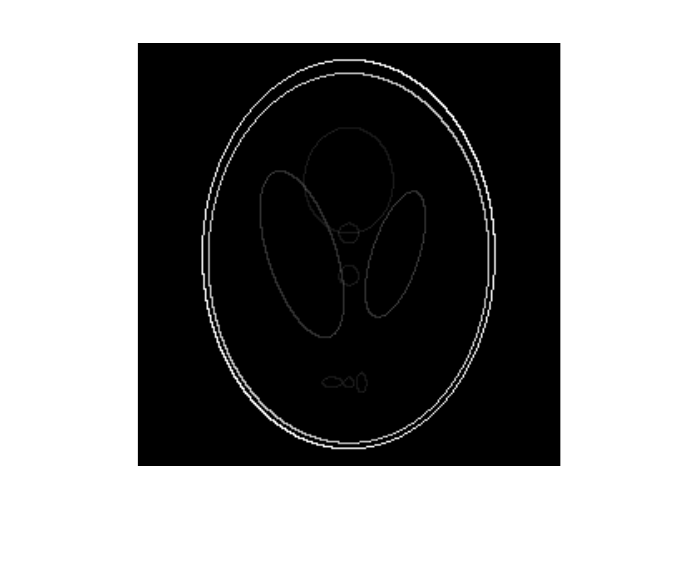}
\end{center}
\caption{The original $256\times 256$ Shepp-Logan phantom  (left), the Shepp-Logan phantom and the magnitudes of its gradient with sparsity $s=2184$ (Fannjiang 2013. Reprinted with permission).}
\label{fig2'}
\end{figure}

 \begin{figure}[t]
\begin{center}
\includegraphics[width=0.5\textwidth]{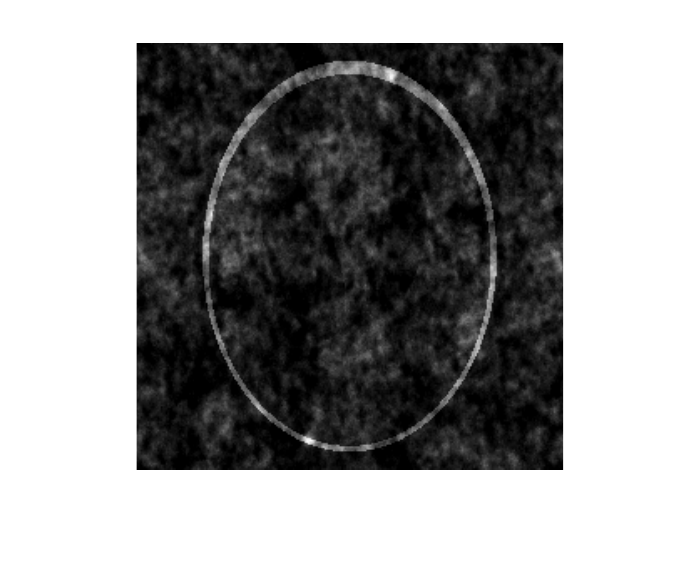}\hspace{-1.5cm}
\includegraphics[width=0.5\textwidth]{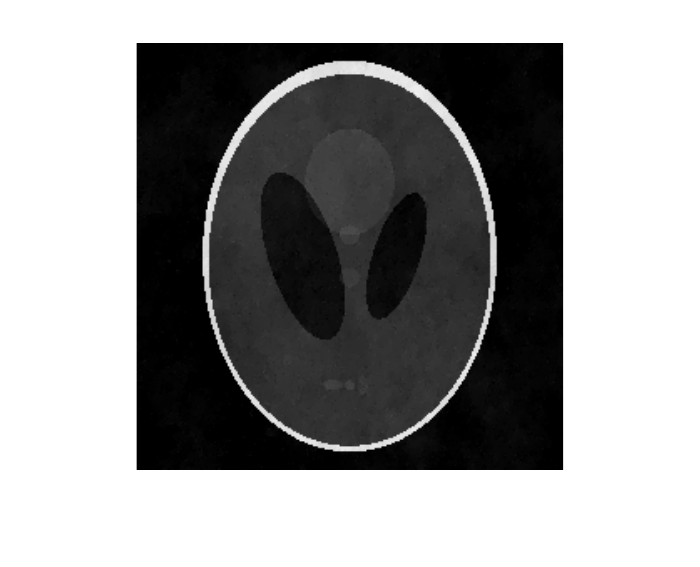}
\commentout{
\includegraphics[width=0.3\textwidth]{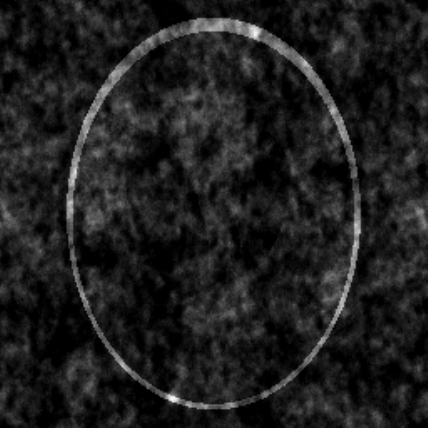}
\includegraphics[width=0.3\textwidth]{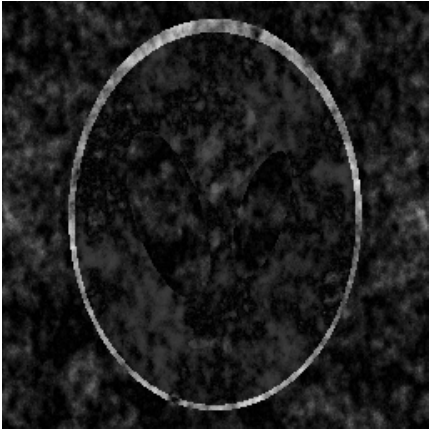}
\includegraphics[width=0.3\textwidth]{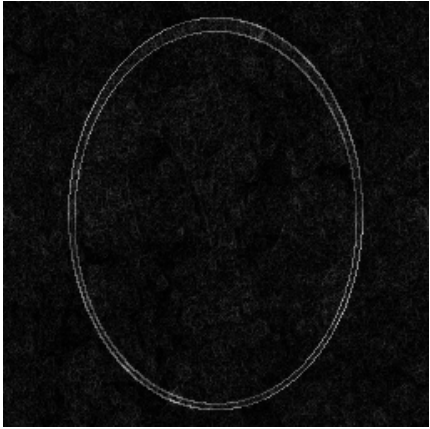}
}
\caption{BPDN reconstruction without external noise (left) and
TV-min reconstruction with $5\%$ noise (right) (Fannjiang 2013. Reprinted with permission).
}
\label{fig3'}
\end{center}
\end{figure}

Fig. \ref{fig2'} and Fig. \ref{fig3'} are a numerical demonstration of TV-min reconstruction of 2D object (the phantom).
Fig. \ref{fig2'} shows the original image and its gradient which is sparse compared to the original dimensionality. Fig. \ref{fig3'} shows
the reconstruction with BPDN (left) and
TV-min (right). TV-min performs well as expected because the TV-sparsity is 
the correct prior for the object.  On the other hand, BPDN performs poorly because
the L1-sparsity is the wrong prior. 

\subsection{BPDN for joint sparsity}\label{sec3.2}
The close relationship between \eqref{tv1} and \eqref{32} can be seen from the following equation for the 1D setting 
\[
(e^{2\pi i \xi_j}{  -1})g_j=\sum_k e^{-2\pi i \xi_j k} (f_{k+1}-f_k).
\]
In other words,  the new data vector $\tilde \bg=((e^{2\pi i \xi_j}-1)g_j)$, 
the new noise vector  $\tilde\be=((e^{2\pi i \xi_j}-1)e_j)$
and the new object vector $\tilde \mbf=(f_{k+1}-f_k)$ are related via the same
sensing matrix as for BPDN. Clearly, $|\tilde e_j|\leq 2|e_j|, j=1,...,M$.
Moreover, if $e_j$ are independently and identically distributed, then $\tilde e_j$ are also independently and identically distributed with variance
\[
\IE|\tilde e_j|^2=\IE|e^{2\pi i \xi_j}-1|^2 \times  \IE|e_j|^2=2  \IE|e_j|^2
\]
when $\xi_j$ is the uniform random variable over $[-1/2,1/2]$.
Hence for large $M$ the new noise magnitude $\|\tilde\be\|_2\approx \sqrt{2}\|\be\|_2.$ Here and below $  \IE$ denotes the expected value

The similar relationship exists in the 2D case.  
Let  $\mbf_j=\Delta_j \mbf$ which satisfy the linear constraint
\beq
\label{28}
\Delta_1 \mbf_2=\Delta_2  \mbf_1. 
\eeq
Define
\beqn
\bg_1&=[(e^{2\pi i \xi_j}-1)g_j],\quad 
\bg_2&= [(e^{2\pi i \eta_j}-1)g_j]\\
\be_1&=[(e^{2\pi i \xi_j}-1)e_j],\quad
\be_2&= [(e^{2\pi i \eta_j}-1)e_j]\\
\eeqn
where $\xi_j,\eta_j, j=1,...,M$ are independent uniform random variables over $[-1/2,1/2]$. 
Then  $\bF=[\mbf_1,\mbf_2]\in   \IC^{N\times 2}$,  
$\bG=[\bg_1,\bg_2]\in   \IC^{M\times 2}$ and $\bE=[\be_1,\be_2]$ are
related through 
\[
\bG=[\bPhi\mbf_1,\bPhi\mbf_2]+\bE 
\]
subject to the linear constraint \eqref{28}.
This formulation calls for the $L^1$-minimization (Fannjiang 2013) 
 \beq
\min \|[\bh_1,\bh_2]\|_{2,1},\quad\hbox{s.t.}
\quad \|\bG-[\bPhi\bh_1,\bPhi\bh_2]\|_{\rm F}\leq \|\bE\|_{\rm F}, 
\label{bp1}
\eeq
subject to the constraint
\beq
\label{31}
\Delta_2\bh_1=\Delta_1\bh_2
\eeq
where $\|\cdot\|_{\rm F}$ is the Frobenius norm and   $\|\cdot\|_{2,1}$ is the  the mixed $(2,1)$-norm (Benedek and Panzone 1961, Kowalski 2009). 
\beq
\|\bX\|_{2,1}&=&\sum_{j}  \|\hbox{\rm row}_j (\bX)\|_2. \label{90}
\eeq
 The reason for minimizing the mixed $(2,1)$-norm in \eqref{bp1} is
 that $\mbf_1$ and $\mbf_2$ share the same sparsity pattern which should be enforced. 
 
 To get a more clear idea about $\|\bE\|_{\rm F}$, we apply the same analysis as above and obtain  
 \[
 \|\be_i\|_2^2\approx \IE\|\be_i\|_2^2=2\IE\|\be\|_2^2,\quad i=1,2, 
 \]
 for sufficiently large $M$.

 The convex program \eqref{bp1}-\eqref{31} is an example of BPDN with constrained joint sparsity. More generally, suppose that the columns of the unknown multi-vectors $\bF\in   \IC^{N\times J}$ share
 the same support and are related to the data multi-vectors 
$\bG\in  \IC^{M\times m}$ and the noise multi-vectors $\bE\in  \IC^{M\times J}$  via 
\beq
\label{mm}
\bG=[\bPhi_1\mbf_1,\bPhi_2\mbf_2,...,\bA_J\mbf_J]+\bE 
\eeq
subject to the linear constraint $\cL \bF=0$. 

For this setting,  the following formulation of  BPDN with joint sparsity
is natural 
\beq
\min \|\bH\|_{2,1},\quad\hbox{s.t.}
\quad \|\bG-[\bPhi_1\bh_1,\bPhi_2\bh_2,...,\bA_J\bh_J]\|_{\rm F}\leq \ep, \quad\hbox{s.t}\quad \cL\bH=0,
\label{jbp}
\eeq
with $\ep =\|\bE\|_{\rm F}$. 

\subsection{OMP for joint sparsity}\label{sec3.3}
Next we present an algorithmic extension of OMP for joint-sparsity 
(Cotter {\em et al.} 2005, Chen and Hua 2006, Tropp {\em et al.} 2006)
 to the setting with  multiple sensing matrices \eqref{mm} (Fannjiang 2013).

\begin{center}
   \begin{tabular}[width=5in]{l}
   \hline
   \centerline{{\bf Algorithm 2.}\quad  OMP for joint sparsity} \\ \hline
   Input: $\{\bA_j\}, \bb,\ep>0$\\
 Initialization:  $\mbx^0 = 0, \bR^0 = \bG$ and $\cS^0=\emptyset$ \\ 
Iteration: {  For $k=1,2,3,\cdots$} \\
\quad 1) $i_{\rm max} = \hbox{arg}\max_{i}\sum^J_{j=1}|\Phi^\dagger_{j,i}R^{k-1}_j |,\hbox{where $\Phi^\dagger_{j,i}$ is the conjugate transpose  of $i$-th column of $\bPhi_j$} $\\
 \quad      2) $\cS^k= \cS^{k-1} \cup \{i_{\rm max}\}$ \\
  \quad  3) $\bF^k = \hbox{arg} \min\|
     [\bA_1\bh_1,...,\bA_J\bh_J]-\bG\|_{\rm F}$ s.t. \hbox{supp}($\bH$) $\subseteq S^k$ \\
  \quad   4) $\bR^k = \bG- [{  \bA_1}\mbf^k_1,...,{ \bA_J}\mbf^k_J]$\\
\quad  5)  Stop if $\sum_j\|R^k_j\|_{2}\leq \ep$.\\
 Output: $\bF^k$. \\
 \hline
   \end{tabular}
\end{center}
\bigskip

Note that the linear constraint $\cL$ is not enforced in Algorithm 2.
The idea is to first find the support of the multi-vectors without
taking into account of the linear constraint, and, in the second stage, follow the support recovery with least squares
\beq
\label{ls2}
\bF_*=\hbox{\rm arg} \,\min_{\bH}{  \|\bG-[\bA_1\bh_1,...,\bA_J\bh_J]\|_{\rm F}},\quad \hbox{s.t.}\quad  \supp (\bH)\subseteq \supp (\bF^\infty),\quad \cL\bH=0
\eeq
where $\bF^\infty$ is the output of Algorithm 2.

  For more discussion and applications of constrained joint sparsity, the reader is referred to Fannjiang 2013a where the performance guarantees similar to Theorem \ref{thm:rip} and Theorem \ref{thm2} 
 are proved for constrained joint sparsity.  

\section{Fresnel diffraction with point objects}\label{sec4}\label{sec:blo}
\commentout{
For simplicity of notation, we choose the physical units such that
$\om/z_0=2\pi$ with which the resolution length (RL) is conveniently normalized to 1.  With this, the paraxial Green function has the
form
\[
{e^{i\om z_0}\over 4\pi z_0}e^{i\pi (x^2+y^2)}e^{-2\pi i (x\xi +y\eta)} e^{i \pi(\xi^2+\eta^2)}
\]

 Let $(x_k,y_k), k=1,...,s$ be the coordinates of the point objects
 on the object plane where the sparsity $s$ is the number of point objects. 
 Let $(\xi_l,\eta_l) \in [0,1]^2, l=1,...,M$ be the coordinates of
 the sensors on the sensor plane. 
 A key idea  of compressive
sensing matrix is to randomize the locations of
the sensors within the given aperture.  So we  assume $\xi_j,\eta_j$ are independent
uniformly distributed in $[0,1]$.

Then the measurement data  $\bg=(g_j)_1^N$at sensor $j$ after multiplying  $e^{-i \pi(\xi_j^2+\eta_j^2)}$
 can be expressed as
 \beq
 \label{eq1}
 g_j=\sum_{k=1}^sf_k e^{-2\pi i (x_k\xi_j+y_k\eta_j)},  \quad j=1,...,M,
 \eeq
 where $f_k, k=1,...,s$ is the strength of the point object $k$ multiplied by the quadratic phase factor $e^{i\pi (x_k^2+y_k^2)}$ and
 the constant factor ${e^{i\om z_0}\over 4\pi z_0}$.

  In the form of \eqref{eq1} the imaging problem of determining  $\mbf=(f_k)_1^s$ and $(x_k,y_k), k=1,...,s$ from $\bg$ is clearly a nonlinear one. The standard linearization scheme is to divide
  the object plane into a grid that is resolvable by the aperture
  $[0,1]^2$ and, under the choice of coordinates above, can be set as 
   $\IZ^2$. 
  }
   
   A major problem with discretizing the object domain shows up when  the objects are point-like. In this case it is  unrealistic to assume the objects
   are located exactly on the grid as the  forceful matching between the point objects and the grid can create detrimental errors.
 Without additional prior information the gridding error due to the mismatch between the point object locations and the grid points can be  as large
as the data themselves, resulting in  a low Signal-to-Noise Ratio (SNR). 

We shall call the grid spacing $\ell$ given in  \eqref{pix} the {\em Resolution Length} (RL), which is the natural unit for resolution analysis. In the RL unit, the object domain grid becomes {  a subset of}
the integer grid $\IZ$.  

In the case of point objects, to refine the standard grid  and reduce discretization error we consider a fractional grid
\beq
\label{23}
\IZ/F=\{ j/F: j\in \IZ\}
\eeq
where $F\in \IN$ is called {\em the
refinement factor}.
The  random partial Fourier  matrix \eqref{20'} now takes the form
  \beq
  \label{86}
 {  \bA}=\lt[ e^{-i2\pi \xi_j k/F}\rt]  \eeq
 where $\xi_j\in [-1/2,1/2]$ are independent uniform random variables.    In the following numerical examples, we shall consider
  both deterministic (see \eqref{87})  as well as random sampling schemes. 
  
\begin{figure}[h]
\begin{center}
\includegraphics[width=8cm]{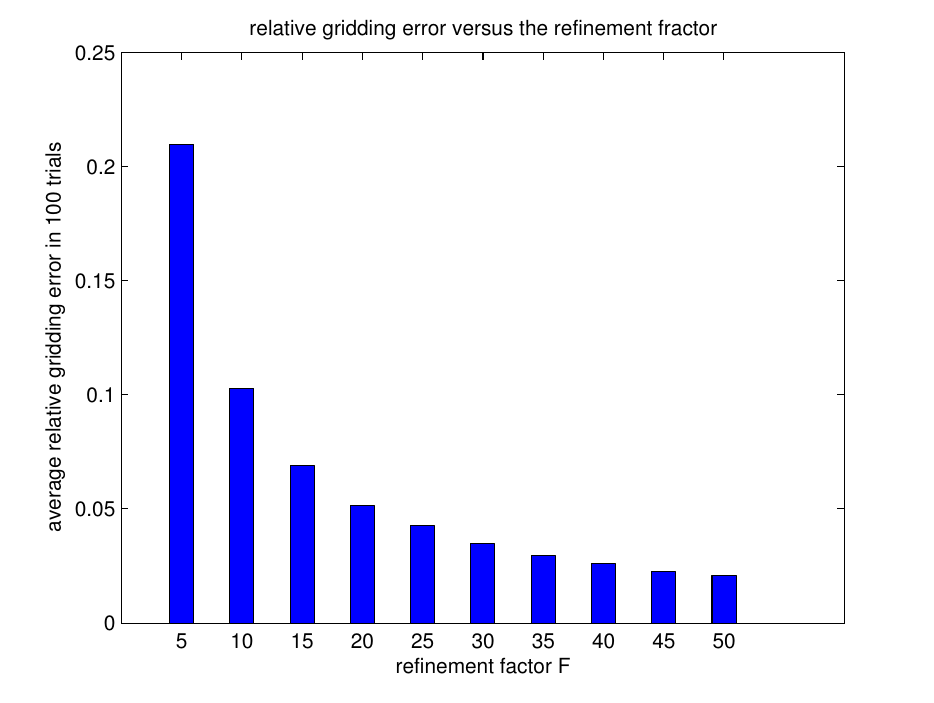}
\caption{ 
The relative gridding error is roughly inversely proportional to the refinement factor. 
(Fannjiang and Liao 2012a. Copyright \copyright 2012 Society for Industrial and Applied Mathematics. Reprinted with permission. All rights reserved)}
\label{fig328-1}
\end{center}
\end{figure}

   \commentout{
\beq\label{1}
    \bPhi \mbf + \mbe = \bg
    \label{linearsystem}
\eeq
 where the error vector 
 $\mbe=(e_k)\in \IC^M$ is the sum of  the external noise
 $\bn=(n_k)$ and the discretization or gridding  error $\bd=(\delta_k) \in \IC^M $
 due to mismatch between the object locations and
 the grid points.
 } 
As shown in Fig. \ref{fig328-1}, the relative gridding error
$\|\bd\|/\|\bPhi\mbf\|$ is roughly inversely proportional to
the refinement factor $F$. 

\begin{figure}[h]
\begin{center}
  \includegraphics[width=8cm]{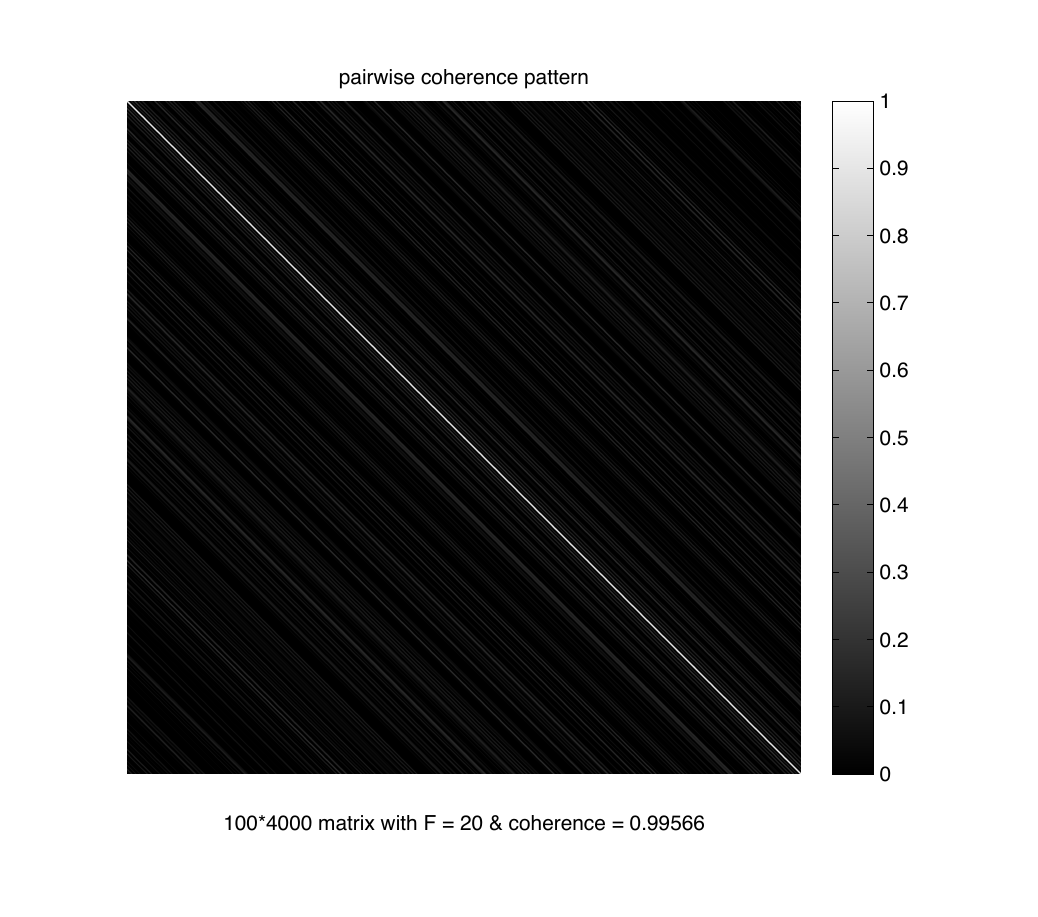}
   \includegraphics[width=8cm,height=7cm]{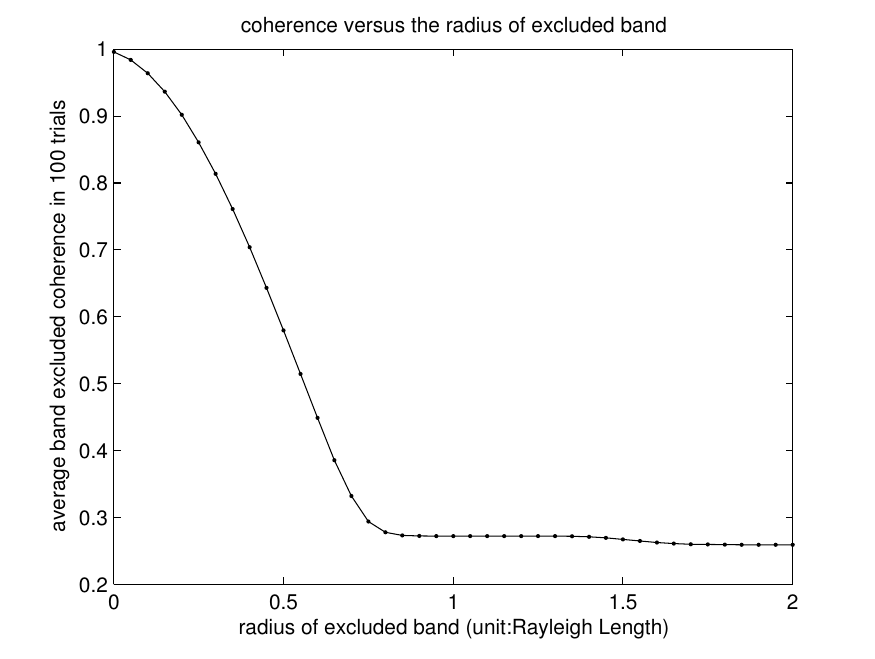}
 \caption{ Coherence pattern $[\mu(j,k)]$ for the $100\times 4000$  matrix with $F=20$ (left).   The off-diagonal elements tend to diminish as the  row number increases. The coherence band near the diagonals, however, persists, and has the average profile shown on the right panel where the vertical axis is the pairwise coherence averaged over 100 independent trials and
 the horizontal axis is the distance between two object points (Fannjiang and Liao 2012a. Copyright \copyright 2012 Society for Industrial and Applied Mathematics. Reprinted with permission. All rights reserved).}
  \label{figurecoherencepattern}\label{fig1}
  \end{center}
\end{figure}

Fig.  \ref{fig1} shows the coherence pattern $[\mu(j,k)]$ of
a $100\times 4000$ matrix (\ref{86}) 
with $F=20$ (left panel). The bright diagonal band represents a heightened correlation (pairwise coherence) between
a column vector and its neighbors  on both sides (about 30). 
The right panel of Figure \ref{fig1} shows  a half cross section of the coherence band across two RL, {  averaged over 100 independent trials}.     
In general sparse recovery with large $F$ exceeds the capability of currently known
algorithms as the condition number of the  $100\times 30$ 
submatrix corresponding to the coherence band in Figure \ref{fig1}
easily exceeds  $10^{15}$. The high condition number makes 
stable recovery  impossible.  
While  Figure \ref{fig1} is typical of
the coherence pattern of one-dimensional sensing matrices,  the coherence pattern  for two or three dimensions  is considerably more complicated depending on how the objects are vectorized.

\subsection{BLOOMP}\label{sec4.1}
To overcome the conundrum of highly coherent sensing matrix due to a refined grid,
we have to go beyond the coherence parameter and study the coherence pattern of
the sensing matrix. 

 The coherence pattern of a sensing matrix can be described in terms of the notion of {\em coherence band} defined below.
 Let $\eta>0$. Define the $\eta$-coherence band  of the index $k$ as \begin{equation}
   B_\eta(k) = \{i\ | \ \mu(i,k) > \eta\},
   \label{singleneighbor}
\end{equation}
and   the {  double} coherence band as
\beq
   B^{(2)}_\eta(k) &\equiv& B_\eta(B_\eta(k))= \displaystyle \cup_{j\in B_\eta(k)} B_\eta(j)
\eeq

The first technique for taking  advantage of the
prior information of well separated objects  is called Band Exclusion (BE) and 
can be easily embedded in  the greedy algorithm,
Orthogonal Matching Pursuit (OMP). 

To imbed BE into OMP, 
we make the following change to  the matching step 
\[
i_{\rm max} = \hbox{arg}\min_{i}{  |\lan \br^{n-1},\Phi_i\ran | }, \quad i \notin B^{(2)}_\eta(S^{n-1}),\quad n=1,2,....
\]
meaning that  the double $\eta$-band of the estimated  support  in the previous iteration is avoided in the current search. This is natural if the sparsity pattern
of the object is such that
$B_\eta(j), j\in \hbox{supp}(\mbx)$ are pairwise disjoint. 
We call the modified algorithm  the Band-excluded Orthogonal Matching Pursuit (BOMP) as stated in  { Algorithm 3}.

\begin{center}
   \begin{tabular}{l}
   \hline
   \centerline{{\bf Algorithm 3.}\quad Band-Excluded Orthogonal Matching Pursuit (BOMP)} \\ \hline
   Input: $\bA, \bb,\eta>0$\\
 Initialization:  $\mbx^0 = 0, \br^0 = \bb$ and $S^0=\emptyset$ \\ 
Iteration: For  $j=1,...,s$\\
\quad {1) $i_{\rm max} = \hbox{arg}\max_{i}{  |\lan \br^{j-1},\Phi_i\ran | }, i \notin B^{(2)}_\eta(S^{j-1}) $} \\
  \quad      2) $S^{j} = S^{j-1} \cup \{i_{\rm max}\}$ \\
  \quad  3) $\mbx^j = \hbox{arg} \min_\bh \|
     \bA \bh-\bb\|_2$ s.t. \hbox{supp}($\bh$) ${ \subseteq} S^j$ \\
  \quad   4) $\br^j = \bb- \bA \mbx^j$\\
 Output: $\mbx^s$. \\
 \hline
   \end{tabular}
\end{center}

\bigskip
The following theorem gives 
 a (pessimistic) performance guarantee for BOMP. 
\begin{theorem}
\label{thm1} (Fannjiang and Liao 2012a)
Let $\mbx$ be $s$-sparse. Let $\eta>0$ be fixed. 
Suppose that
\beq
\label{sep}
B_\eta(i)\cap B^{(2)}_\eta(j)=\emptyset, \quad \forall i, j\in \hbox{supp}(\mbx)
\eeq
and that
    \beq
    \eta(5s -4)\frac{\xmax}{\xmin} + \frac{5\|\mbe\|_2}{2\xmin} < 1 
    \label{RMIP}
    \eeq
    where
    \[
    \xmax = \max_{k} |f_k|,\quad  \xmin = \min_{k} |f_k|.
    \]
    Let $\mbx^s$ be the BOMP reconstruction. 
Then $\hbox{supp}(\mbx^s)\subseteq B_\eta(\hbox{supp}(\mbx))$
and moreover every nonzero component of $\mbx^s$ is in
the $\eta$-coherence band of a unique nonzero component of $\mbx$. 
  \label{momp}
\end{theorem}
\commentout{
\begin{remark}
\label{rmk1'}

In the case of the matrix 
(\ref{3}),  if every two indices in $\suppx$ is more than
one RL apart, then $\eta$ is small for sufficiently
large $N$, cf. Figure \ref{fig1}. 
  
 When the dynamic range ${\xmax}/{\xmin} = \cO(1)$, 
 Theorem \ref{thm1} guarantees approximate recovery
 of  $\cO(\eta^{-1})$ sparsity pattern by BOMP.

\end{remark}
}
\begin{remark}

Condition  (\ref{sep}) means 
that BOMP guarantees to resolve 3 RL.
In practice, BOMP can resolve 
objects separated by close to 1  RL  when the dynamic range is nearly 1.  

\end{remark}
\begin{remark}\label{rmk2'} A main difference between  Theorem \ref{thm2} and 
Theorem $\ref{thm1}$
 lies in the role
played by  the dynamic range $\xmax/\xmin$
and the separation condition (\ref{sep}).

Another difference is approximate recovery of support in Theorem \ref{thm1} versus exact recovery of support in Theorem \ref{thm2} (a). 
In contrast to $F$-independent nature of approximate support recovery, exact support recovery would probably be highly sensitive to
the refinement factor $F$. That is, as  $F$ increases, 
the chance of missing some points in the support set also increases.
As a result, the error of reconstruction $  \|f^s-f\|_2$ tends to increase
with $F$ (as evident in Fig. \ref{fig3}). 

\end{remark}

A main shortcoming  with
BOMP is in its failure to perform even when the dynamic range is  even moderately greater than unity. 
To overcome this problem, we introduce 
the second technique: 
the {\em Local Optimization} (LO) 
which is a residual-reduction technique  applied
to the current estimate $S^k$ of the object support (Fannjiang and Liao 2012a).  

\begin{center}
   \begin{tabular}{l}
   \hline  
   \centerline{{\bf Algorithm 4.}\quad  Local Optimization (LO)}  \\ \hline
    Input:$\bA,\bb, \eta>0,  S^0=\{i_1,\ldots,i_k\}$.\\
Iteration:  For $j=1,2,...,k$.\\
\quad 1) $\mbx^j= \hbox{arg}\,\,\min_{\bh}\|\bA \bh-\bb\|_2,\quad
 \hbox{supp}(\bh)=(S^{j-1} \backslash \{i_j\})\cup \{i'_j\}, $  $  i'_j\in B_\eta(\{i_j\})$.
\\
 \quad 2) $S^j=\hbox{supp}(\mbx^j)$.\\
    Output:  $S^k$.\\
    \hline
   \end{tabular}
\end{center}


In other words, given a support estimate $S^0$, LO fine-tunes 
the support estimate by 
adjusting each element in $S^0$ 
within its coherence band in order to
minimize the residual.
The  object amplitudes for the improved support estimate  are obtained by solving the least squares problem. Because of
the local nature of LO, the computation is efficient.

Embedding LO in BOMP gives rise to the Band-excluded, Locally
Optimized Orthogonal Matching Pursuit (BLOOMP).

\begin{center}
   \begin{tabular}{l}
   \hline
   \centerline{{\bf Algorithm 5.} Band-excluded, Locally Optimized Orthogonal Matching Pursuit (BLOOMP)} \\ \hline
   Input: $\bA, \bb,\eta>0$\\
 Initialization:  $\mbx^0 = 0, \br^0 = \bb$ and $S^0=\emptyset$ \\ 
Iteration: For  $j=1,...,s$\\
\quad {1)  $i_{\rm max} = \hbox{arg}\max_{i}|\lan \br^{j-1},\Phi_i\ran | , i \notin B^{(2)}_\eta(S^{j-1}) $} \\
  \quad      2) $S^{j} = \hbox{LO}(S^{j-1} \cup \{i_{\rm max}\})$~where~{  $\hbox{LO} (S^{j-1} \cup \{i_{\rm max}\})$  is the output of Algorithm 4} \\
\hspace{1cm}{  with $S^{j-1} \cup \{i_{\rm max}\}$ as input}.\\
  \quad  3) $\mbx^j = \hbox{arg}  \min_\bh \|
     \bA \bh-\bb\|_2$ s.t. \hbox{supp}($\bh$) $\in S^j$ \\
  \quad   4) $\br^j = \bb- \bA \mbx^j$\\
 Output: $\mbx^s$. \\
 \hline
   \end{tabular}
\end{center}

\bigskip

The same BLO technique can be used to enhance the other
well known iterative schemes such as 
SP, CoSaMP (Needell and Tropp 2009), Compressed  Iterative Hard Thresholding (IHT) (Blumensath and  Davies 2009, Blumensath and  Davies 2010)
and the  resulting algorithms are denoted by
BLOSP, BLOCoSaMP and BLOIHT,  respectively, in the numerical results below. We refer the reader to Fannjiang and Liao 2012a for the details and
descriptions of these algorithms. 

{  MATLAB code of Algorithm 3.5 is available on-line at 

{\tt https://www.math.ucdavis.edu/\~\,fannjiang/home/codes/BLOOMPcode.}
}
\subsection{Band-excluding  thresholding}\label{sec4.2}
A related technique that can be used to
enhance BPDN/Lasso for off-grid objects
is  called the 
the  Band-excluding, Locally Optimized Thresholding (BLOT).  

\begin{center}
   \begin{tabular}{l}
   \hline
   
   \centerline{{\bf Algorithm 6.}\quad Band-excluding, Locally Optimized  Thresholding (BLOT)}  \\ \hline
    Input: $\mbx=(f_1,\ldots, f_N)$, $\bA, \bb, \eta>0$.\\
    Initialization:  $S^0=\emptyset$.\\
Iteration:  For $j=1,2,...,s$.\\
\quad 1) $i_j= \hbox{arg}\,\,\max |f_k|, k\not\in B^{(2)}_\eta(S^{j-1}) $.\\
 \quad 2)  $S^j=S^{j-1}\cup\{i_j\}$.\\
    Output: $\mbx^s=\hbox{arg}\min \|\bA\bh-\bb\|_2$, $\hbox{supp}(\bh)\subseteq \hbox{LO}(S^s)$ where $\hbox{LO}$ is the output of Algorithm 4.\\
     \hline
   \end{tabular}
\end{center}

\medskip

\commentout{
The technique BLOT can be used to enhance the recovery capability with unresolved grids  of 
 the $L^1$-minimization principles, Basis Pursuit (BP)
\beq
\min_{\bh}  \|\bh\|_1,\quad\hbox{subject to}\quad 
\bb=\bA \bh.
\eeq
 and the Lasso
 \beq
\min_{\bh} {1\over 2} \|\bb-\bA \bh\|_2^2+\lambda \sigma \|\bh\|_1,
\eeq
where $\sigma$ is the standard deviation of
the each noise component and $\lambda$ is the regularization parameter.  
}
\subsection{Numerical examples}\label{sec4.3}

\begin{figure}[t]
  \centering
  \subfigure[OMP]{
    \includegraphics[ width = 8cm]{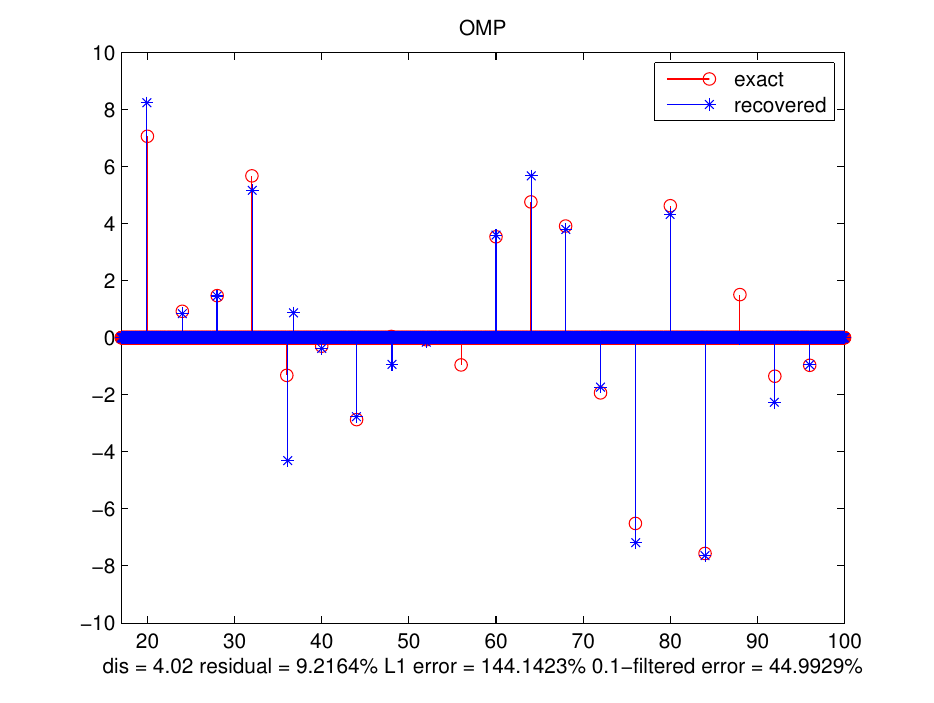}}
          \subfigure[BLOOMP]{
             \includegraphics[width = 8cm]{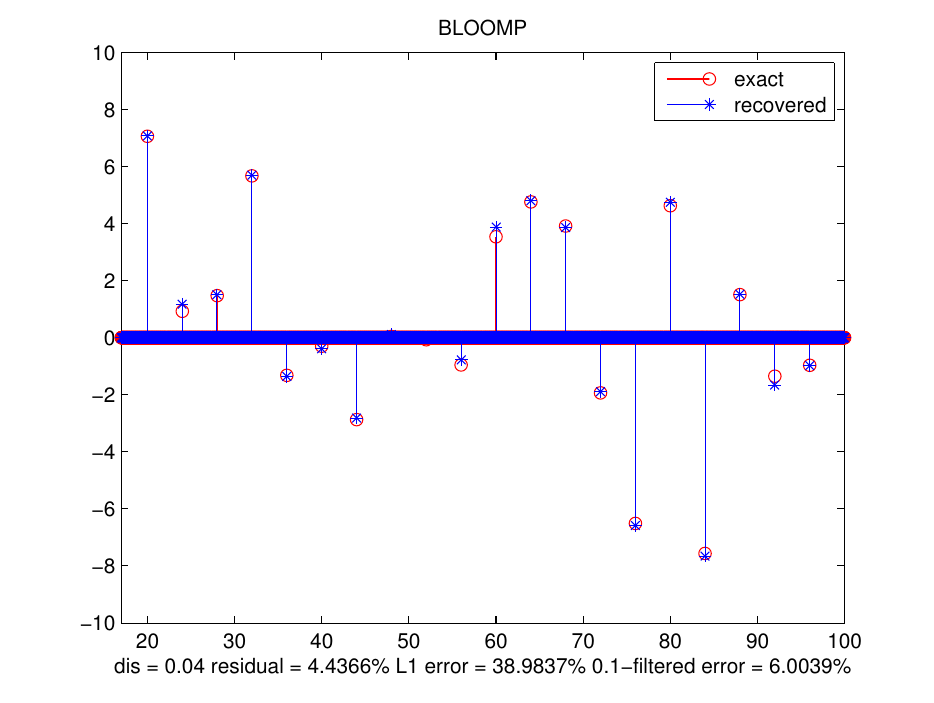}}
          \subfigure[BPDN]{
    \includegraphics[width = 8cm]{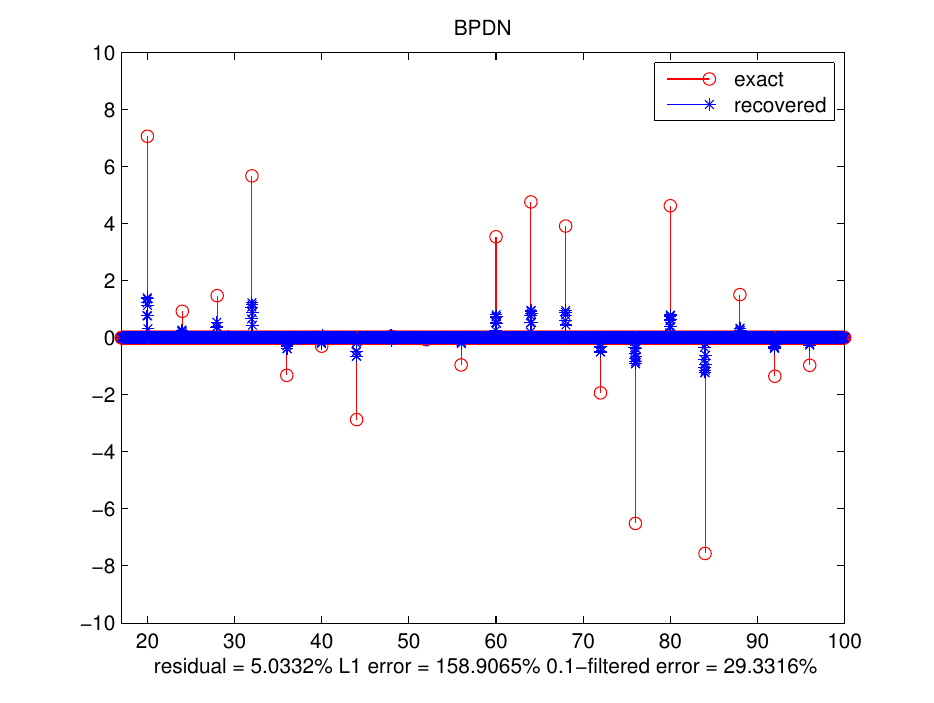}}
          \subfigure[BP-BLOT]{
             \includegraphics[ width = 8cm]{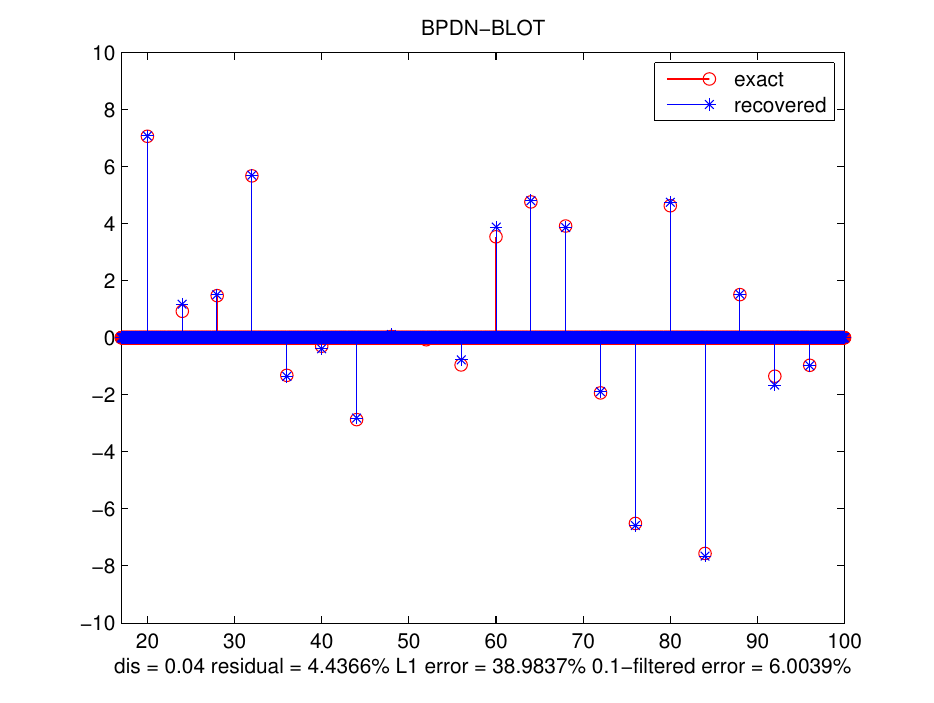}}
          \caption{ Reconstruction by (a) OMP, (b) BLOOMP, (c) BPDN and (d) BPDN-BLOT of the real part of 20 randomly phased spikes with $F=50, \hbox{\rm SNR} = 20$  (Fannjiang and Liao 2012b. Reprinted with permission). }
          \commentout{
          (a) OMP, residual $\approx 7\%$, unfiltered error $\approx 97\%$, $0.1$-filtered error $\approx 28\%$,
           (b) BLOOMP, residual $\approx 4\%$, unfiltered error $\approx 41\%$, $0.1$-filtered error $\approx 6\%$,
            (c) BPDN, residual $\approx 5\%$, unfiltered error $\approx 153\%$, $0.1$-filtered error $\approx 26\%$,
             (d) BPDN-BLOT, residual $\approx 4\%$, unfiltered error $\approx  41\%$, $0.1$-filtered error $\approx 6\%$.  }    
        \label{fig:R1}
\end{figure}

\begin{figure}[t]
  \centering
             \subfigure[SNR=100, $\eta=0$]{
             \includegraphics[width = 5cm]{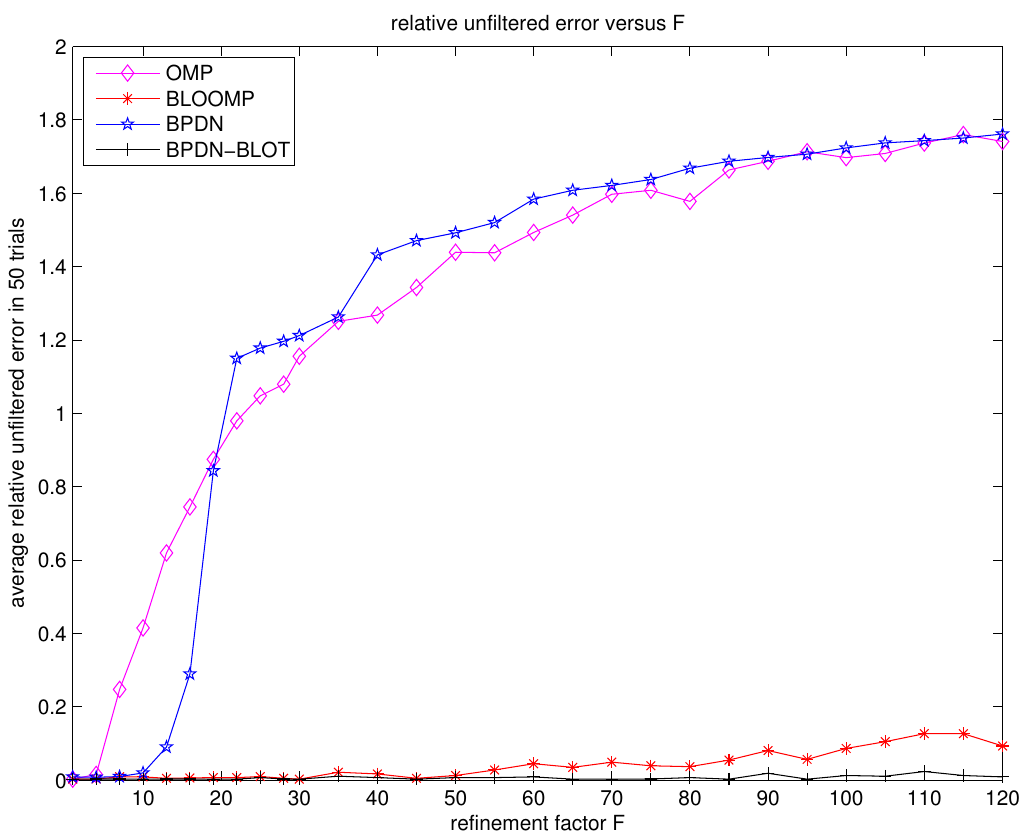}}
                    \subfigure[SNR=20, $\eta=0$]{
             \includegraphics[width = 5cm]{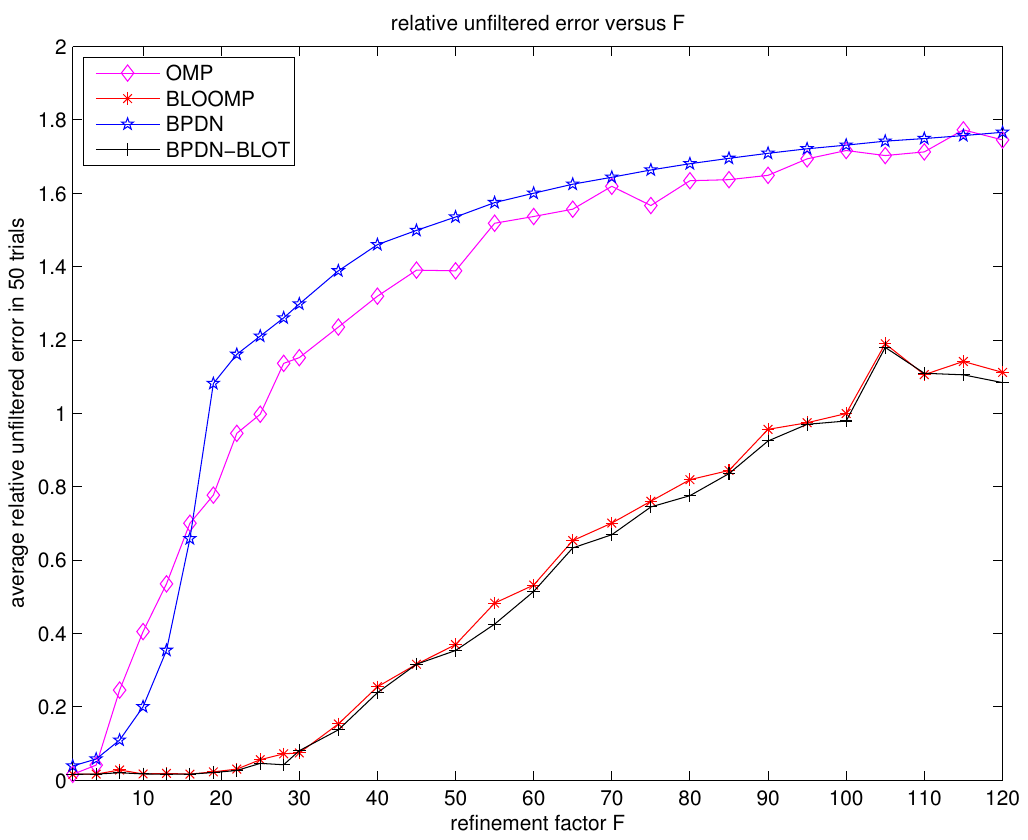}}
                        \subfigure[SNR=10, $\eta=0$]{
             \includegraphics[width = 5cm]{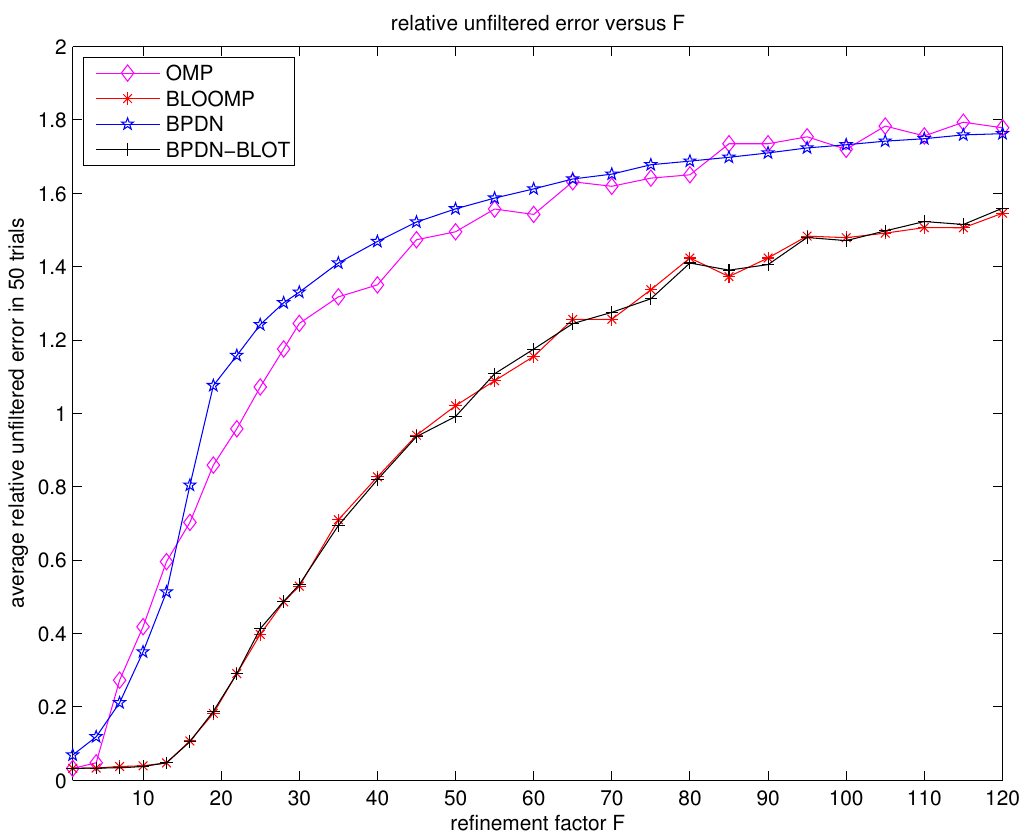}}
          \subfigure[SNR=100, $\eta=0.05\ell$]{
             \includegraphics[width = 5cm]{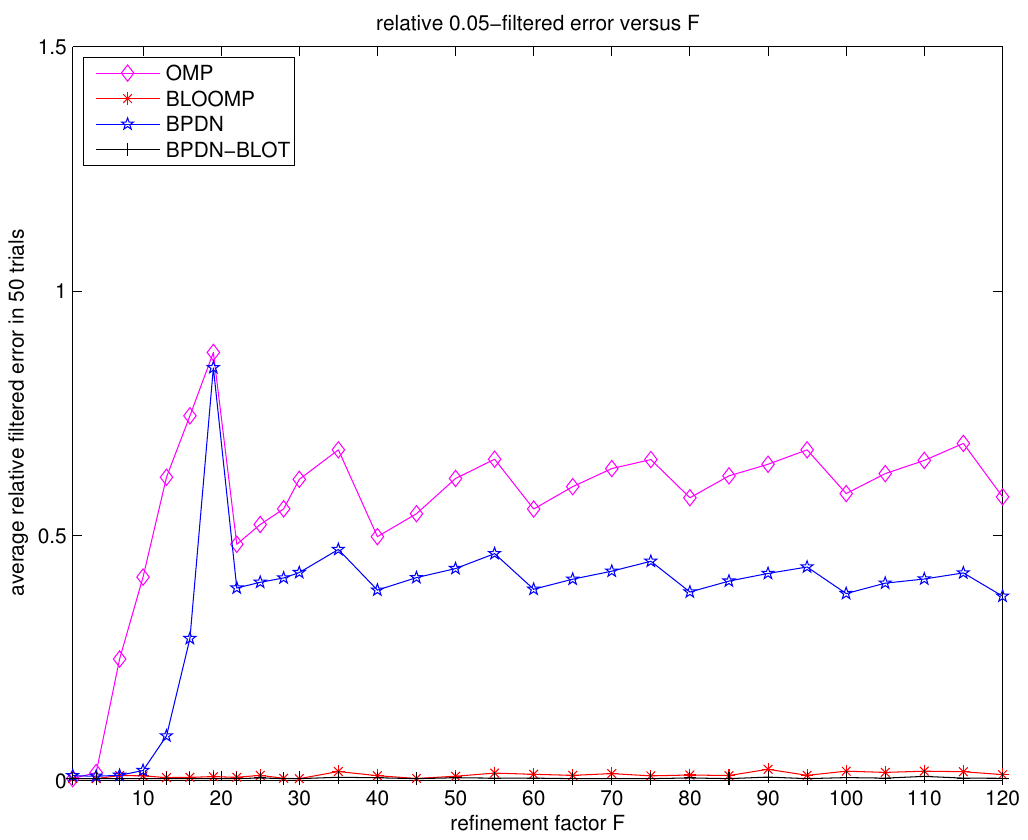}}
                     \subfigure[SNR=20, $\eta=0.05\ell$]{
             \includegraphics[width = 5cm]{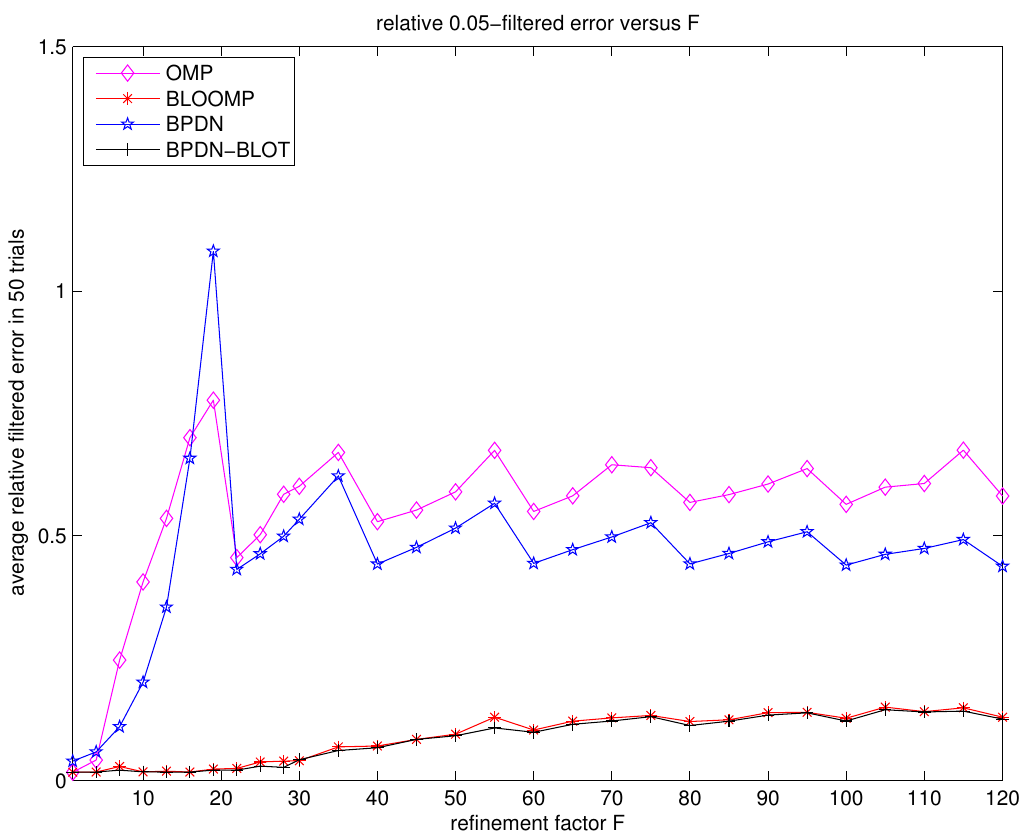}}
        \subfigure[SNR=10, $\eta=0.05\ell$]{
             \includegraphics[width =5cm]{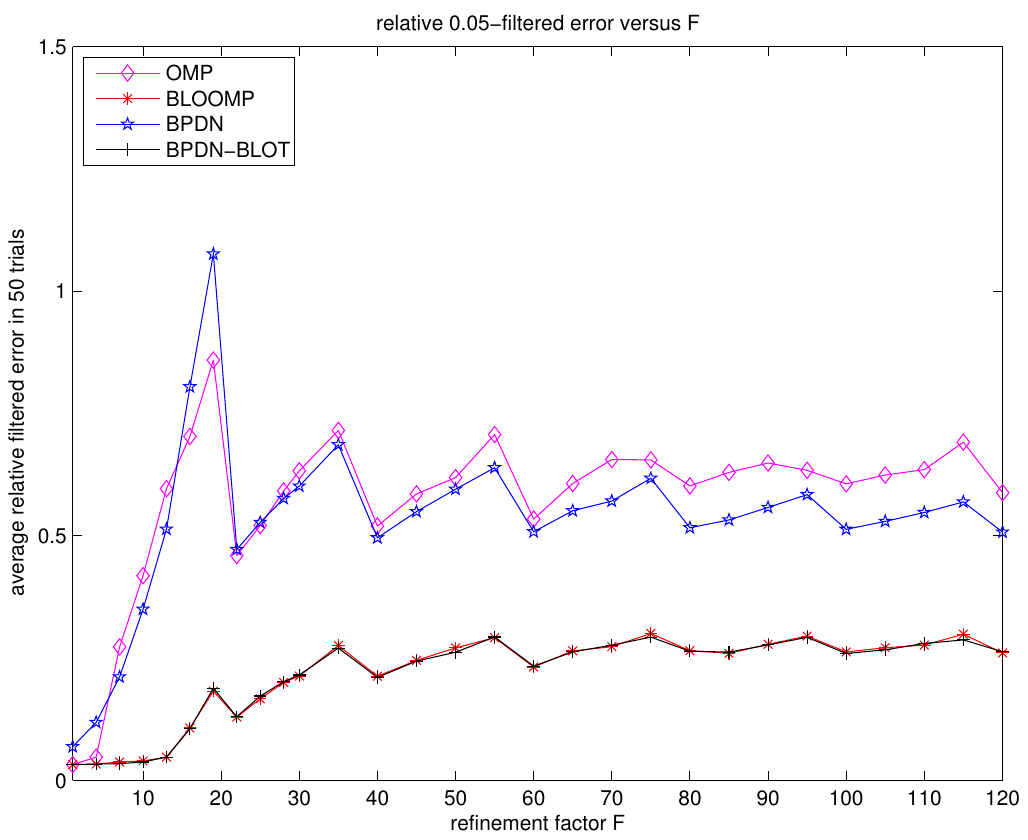}}
              \caption{Relative errors in reconstruction  by OMP, BLOOMP, BP and BP-BLOT as $F$ varies (top) without or (bottom) with filtering (Fannjiang and Liao 2012b. Reprinted with permission).
              }
        \label{fig3}
\end{figure}

For numerical demonstration in Fig. \ref{fig:R1}-\ref{fig3},  we use  {\em deterministic}, 
equally spaced sampling with 
\beq
\label{87}
\xi_j=-{1\over 2}+{j\over M},\quad j=1,...,M
\eeq
and $\bA\in  \IC^{M\times FM}$ with $M=150, F=50$ to recover 20 randomly distributed and  randomly phased point objects (spikes) separated by at least 4 RL.

Fig. \ref{fig:R1} (a)(b)  show how the BLO technique corrects the error of OMP due to the unresolved grid. In particular, several misses are recaptured and false detections removed. 
Fig. \ref{fig:R1} (c) (d) show how the BLOT technique improves
the BPDN estimate. In particular, BLOT has the effect of ``trimming
the bushes" and ``growing the real trees".  
 {  Fig. \ref{fig3} a through c} shows the relative error of reconstruction as a function of $F$ by OMP, BPDN, BLOOMP and BPDN-BLOT  with the same
set-up and three different SNRs. For all SNRs, BLOOMP and BPDN-BLOT produce drastically less  errors compared to OMP and BPDN. 

The growth of relative error with $F$ reflects the sensitivity of the reconstruction error alluded to in Remark
\ref{rmk2'}. 
Note that the  reconstruction error in the {\em discrete} norm can not distinguish how far off the recovered support is from the true object support.  The discrete norm treats any amount of support offset equally. 
An easy remedy to the injudicious treatment of support offset is to use instead 
 the {\em filtered error norm}  
$ 
\| \mbf^s_\eta - \mbf_\eta\|,
$
where $\mbf_\eta$ and $\mbf^s_\eta$ are, respectively,  $\mbf$ and
$\mbf^s$ convoluted with an approximate delta-function of
width $  2\eta$. 

Clearly the filtered error norm is more stable  to
support offset, especially if the offset is less than $\eta$. 
If every spike of $\mbf^s$ is within $\eta$ distance 
from a spike of $\mbf$ {\em and}  if the amplitude differences are  small,
then  the $\eta$-filtered error is small. 
As shown in Fig. \ref{fig3} (d)(e)(f), averaging over $\eta=5\%$ RL produces acceptable filtered error for any refinement factor
relative to the external noise. This suggests that both BPDN-BLOT and BLOOMP recover the object support  on average  within $5\%$ of 1 RL, a significant improvement  over the theoretical guarantee of Theorem \ref{thm1}. 

Next we consider the unresolved partial  Fourier matrix \eqref{86} with
random sampling points to demonstrate the flexibility of the techniques. Let $\xi_j\in [-1/2,1/2], j=1,...,M$ be
independent uniform random variables  with $M=100, N=4000$ and   $F=20$. The test objects are 10 randomly phased and distributed objects, separated by at least 3 RL.
As in Theorem \ref{thm1}, a recovery is counted as a success  if every reconstructed  object
is within 1 RL  of 
the object support. 

\begin{figure}[t]
\begin{center}
\includegraphics[width=8cm]{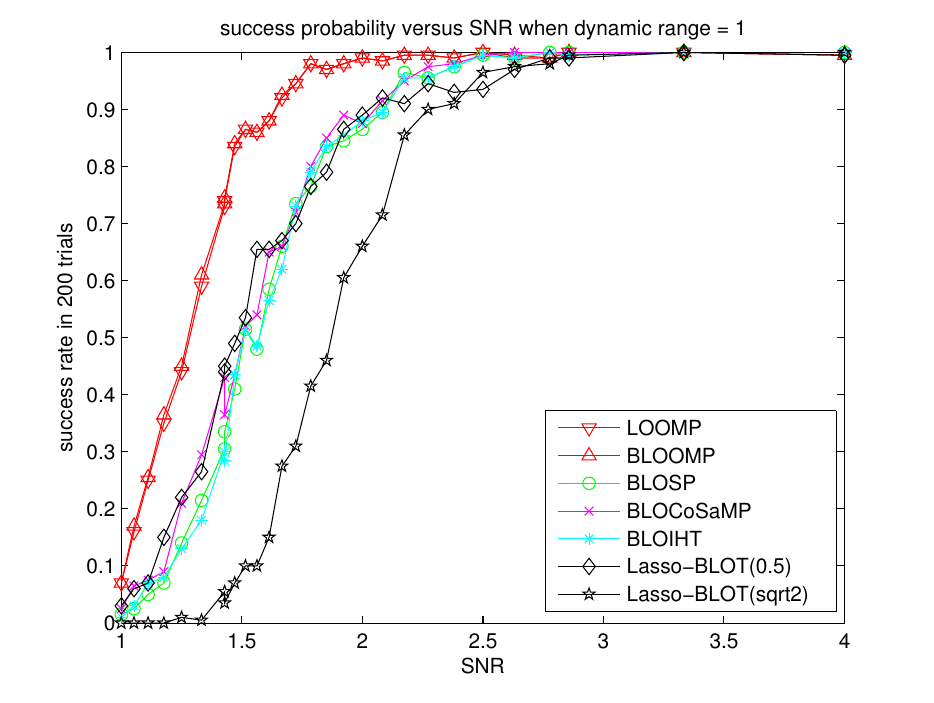}
\includegraphics[width=8cm]{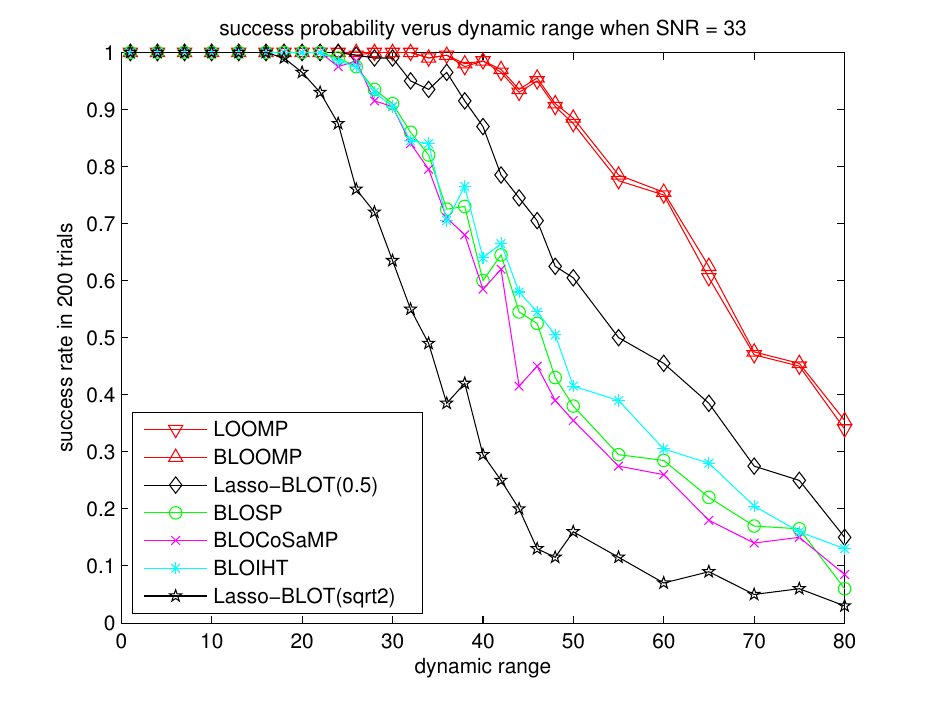}
\caption{Success probability versus (left) SNR  for 
dynamic range 1  and (right) dynamic range for  SNR = 33. Here
LOOMP is a simplified version of BLOOMP and has  nearly identical performance curves (Fannjiang and Liao 2012a. Copyright \copyright 2012 Society for Industrial and Applied Mathematics. Reprinted with permission. All rights reserved).}
\label{fig49-1}
\end{center}
\end{figure}

Fig. \ref{fig49-1} compares  the success rates (averaged over 200 trials)
 of the BLO-enhanced
schemes (BLOOMP, BLOSP, BLOCoSaMP, BLOIHT) 
and BLOT-enhanced scheme (Lasso-BLOT).
Lasso-BLOT
is implemented with the regularization parameter
\beq
\label{10.3}
\lambda=0.5\sqrt{\log{N}}\quad \hbox{\rm (black curves with diamonds)} 
\eeq
or
\beq
\label{10.4}
\lambda=\sqrt{2\log{N}} \quad  \hbox{\rm (black curves with stars)} 
\eeq
(Chen {\em et al.} 2001). 
The empirically optimal choice \eqref{10.3} (labelled
as Lasso-BLOT (0.5)) has a much improved performance over the choice \eqref{10.4}. Clearly, BLOOMP is the best performer in noise
stability and dynamic range among all tested algorithms.

\subsection{Highly redundant dictionaries}\label{sec4.4}

Our discussion in Section \ref{sec4} so far is limited to point-like objects.
But  the methods presented 
above are also applicable to a wide variety of cases where the objects
have sparse representations by redundant dictionaries, instead of
orthogonal bases. 

Suppose that the object is sparse in 
a highly redundant dictionary, which by definition, tends to represent
an object by fewer number of elements than a non-redundant one does. For example,
one can combine different orthogonal bases into a dictionary that can sparsify a wider class of objects than any individual base can. On the other hand,  a redundant dictionary tends to produce a larger coherence parameter and be ill suited for compressive sensing. This is the same kind of conundrum about off-grid point-like objects. 

\begin{figure}[t!]
\centering
\includegraphics[width=12cm]{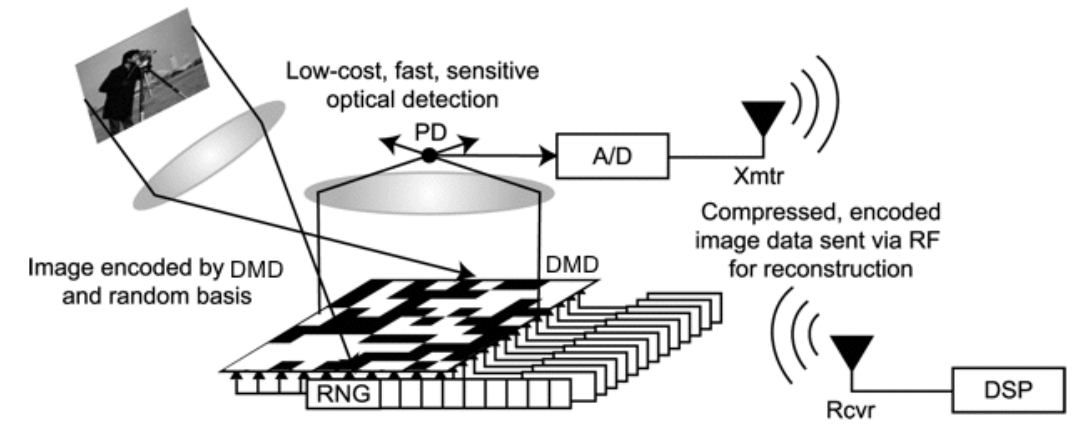}
\caption{Single-pixel camera block diagram ({\tt http://www.dsp.ece.rice.edu/cscamera/})}
\label{fig:cscam}
\end{figure}

One of the most celebrated examples of optical compressive sensing is the {  Single-Pixel Camera} (SPC)  depicted in  Fig. \ref{fig:cscam}.  In SPC,  measurement diversity comes entirely from 
the  Digital Micromirror Device (DMD) instead of sensor array. The DMD consists of an array of electrostatically actuated micro-mirrors. Each mirror can be positioned in one of two states ($\pm 12^\circ$). Light reflected from mirrors in the $+12^\circ$-state only is then collected and focused by the lens and subsequently detected by a single optical sensor. 
For each and every measurement, the  DMD is randomly and independently reconfigured. 
The resulting measurement matrix $ \mathbf{A}$ has  independently and identically distributed entries. 

\begin{figure}[t]
\centering 
\includegraphics[width=8cm]{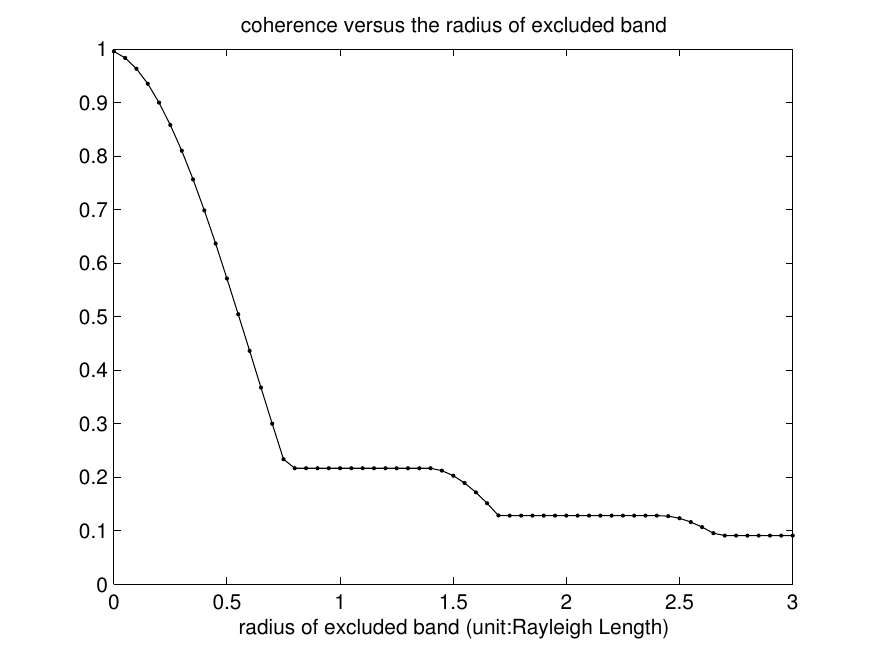}
\includegraphics[width=8cm]{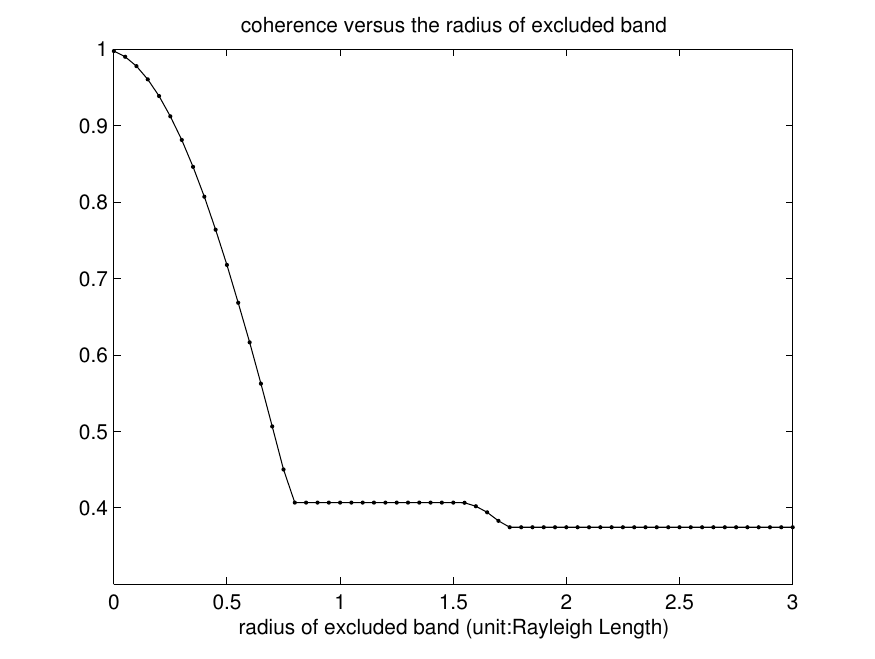}
\caption{The coherence bands of the redundant Fourier frame $\bPsi$ (left) and $\bA=\mathbf{A}\bPsi$ (right), the latter being averaged over
100  realizations of $\mb{A}$ (Fannjiang and Liao 2012a. Copyright \copyright 2012 Society for Industrial and Applied Mathematics. Reprinted with permission. All rights reserved). 
}
\label{fig10}
\end{figure}

Suppose that the object is sparse in terms of a highly redundant dictionary. For simplicity of presentation, consider an 1D object  sparse in an over-complete Fourier frame 
(i.e. a dictionary that satisfies the frame bounds  Daubechies (1992) )
with entries
\beq
\label{102}
\Psi_{k,j} = \frac{1}{\sqrt{R}} e^{-2\pi i \frac{(k-1)(j-1)}{RF}},\quad k=1,...,R,\quad j=1,...,RF,
\eeq
 that includes harmonic as well as non-harmonic modes as its columns, where $F$ is the redundant factor and $R$ is a large integer. In other words, the object can be written as
$\bPsi\mbf$ with a sufficiently sparse vector $\mbf$.
The final  sensing matrix then becomes
\beq
\label{101}
\bA= \mathbf{A}\bPsi.
\eeq 

 The  coherence bands of $\bPsi$ and $\bA$ are  shown in Figure \ref{fig10} from which we see that like Fig. \ref{fig1} the coherence radius is less than 1 RL.
 The same {  BLO- and BLOT}-based  techniques can be applied to \eqref{101}, see Fannjiang and Liao 2012a for numerical results and performance comparison with other techniques for off-grid objects 
 (Cand\`es {\em et al.} 2011, Cand\`es and Fernandez-Granda 2013, Cand\`es and Fernandez-Granda 2014, Duarte and Baraniuk 2013, Tang {\em et al.} 2013). 


 \section{Fresnel diffraction with Littlewood-Paley basis}\label{sec5}

Opposite to the localized pixel basis, 
the Littlewood-Paley basis is  slowly decaying, nonlocal modes based on the wavelet function 
 \beq
 \label{lp2}
 \psi(x)=(\pi x)^{-1} (\sin{(2\pi x)}-\sin{(\pi x)}) 
 \eeq
 which has a compactly supported Fourier transform
  \beq
 \label{lp}
 \hat \psi(\xi)=\int \psi(x) e^{-i2\pi \xi x}d x=\lt\{
 \begin{matrix}
 1, & {1\over 2} \leq |\xi|\leq 1\\
 0,&\hbox{otherwise}.
 \end{matrix}
 \rt.
 \eeq
The following functions 
 \beq
 \psi_{p,q}(x)=2^{-p/2}\psi(2^{-p}x-q),\quad p,q\in \IZ
 \eeq
 form an orthonormal wavelet basis in $L^2(\IR)$ (Daubechies 1992). 
  Expanding  the masked object  $V$ \eqref{mo} in the Littlewood-Paley basis  we write
 \beq
 \label{30}
V(x)=\sum_{p,q\in \IZ}V_{p,q} \psi_{p,q}(x).
 \eeq

The main point of the subsequent discussion is to design a sampling scheme such that the resulting sensing matrix has desirable 
compressive sensing properties (Fannjiang 2009). 

Let $\{2^p: p=- p_*, -p_*+1,..., {  p_*}\}$ be the dyadic scales present in \eqref{30}, $\{q:|q|\leq N_p\}$ the modes present on the scale
$2^p$ and $2M_p+1$ the number of measurements corresponding to
the scale $2^p$. Let
\beq
k&=&\sum_{j=-p_*}^{p'-1}(2M_j+1)+q',\quad |q'|\leq {  M_{p'}},
\quad |p'|\leq p_*\label{47'}
\eeq
be the index for the sampling points. Throughout this section, $k$ is determined by $p', q'$ by \eqref{47'}. Let $x_k$ be the sampling points  and set {  the normalized coordinates}\beq
\label{49'}
{x_k \om \ell\over 2\pi z_0}=\xi_k, \quad  k=1,...M
\eeq
where, as shown below, $\ell$ is a resolution {  length} and  $\xi_k\in [-1/2, 1/2]$ are determined below, c.f. \eqref{xi}. This means that  the aperture (i.e. the sampling range of $x_k$) is again
given by \eqref{27'}.

Let $\bg=(g_k)$ be the data vector with 
\beqn
g_k&=&C^{-1}{u^{\rm s}(x_k,z_0)} e^{-i\om x_k^2/(2z_0)}.
\eeqn
Direct calculation with \eqref{diffract} and \eqref{49'} then gives
\beq g_k
&=& \sum_{p,q\in \IZ} 2^{p/2} V_{p,q} e^{-i2\pi\xi_k \ell^{-1}
2^{p} q}\hat\psi(\xi_k\ell^{-1} 2^{p}),\quad k=1,...,M. \label{7}
\eeq
Let  $\mbf=(f_l)$ be the object vector  with
\[
f_l=(-1)^q{ 2^{p/2} V_{p,q}}
\]
where the indices are related by 
\[ l=\sum_{j=-p_*}^{p-1}(2N_j+1)+q. 
\]

Suppose that  \beq
\label{freq}
\ell {  \le} 2^{-p_*-1} 
 \eeq
i.e. $2\ell$ is {  less} than or equal to the smallest scale
 in the wavelet presentation \eqref{30}. 
 

 Let $\zeta_{p',q'}$ be independent, uniform random variables on 
$[-1/2,1/2]$ and let
\beq\label{50'}
\xi_k={\ell \over 2^{p'}}
\cdot\lt\{\begin{matrix}
1/2+\zeta_{p',q'},& \zeta_{p',q'}\in [0,1/2]\\
-1/2+\zeta_{p',q'},&\zeta_{p',q'}\in [-1/2,0]
\end{matrix}\rt.
\eeq
where $k$ is determined by \eqref{47'}.
  By the assumption (\ref{freq}), we have 
  \[
    \xi_k\in [-1/2,1/2],\quad  \forall  p'\geq -p_*. 
   \]
More specifically, by \eqref{49'}, we have 
\[
x_k\in {2\pi z_0\over \om 2^{p'}}\lt(\lt[-1, -{1\over 2}\rt]\cup \lt[{1\over 2}, 1\rt]\rt),
\]
i.e. the sampling regions for different dyadic scales {  indexed by $p'$} are disjoint with the ones for the smaller scales on the outer skirt of the aperture, taking up a 
bigger portion of the aperture. The resulting sampling points are geometrically concentrated  near (but not exactly at) the center of the aperture.

Let
 the sensing matrix elements be
\beq
\label{sense22}
\Phi_{k,l}= (-1)^q\hat\psi(\xi_k 2^{p}\ell^{-1}) e^{-i2\pi\xi_k
2^{p} q/\ell}. 
\eeq

We claim  that $  \Phi_{k,l}=0$ for $p\neq p'$. This is evident
from \eqref{50'} and  the following calculation
\beq
\label{52'}
{  \ell^{-1}}\xi_k 2^p={2^{p-p'}}
\cdot\lt\{\begin{matrix}
1/2+\zeta_{p',q'},& \zeta_{p',q'}\in [0,1/2]\\
-1/2+\zeta_{p',q'},&\zeta_{p',q'}\in [-1/2,0].
\end{matrix}\rt.
\eeq
For $p\neq p'$  the absolute value of \eqref{52'} is  either greater than 1 or
less than 1/2  and hence \eqref{52'} is outside  
the support of $\hat \psi$ . 

On the other hand, for $p=p'$,  \eqref{52'} is inside the support of $\hat \psi$ and so 
\beq
\label{sense2}
\Phi_{k,l}=e^{-i 2 \pi q\zeta_{p,q'}},\quad
|q'|\leq M_{p},\quad |q|\leq N_p
\eeq
which constitute  the same random partial Fourier matrix that we have seen
above. 
In other words, under the assumption (\ref{freq}) 
the sensing matrix $\bPhi=[\Phi_{k,l}]\in \IC^{M\times N}$,
with  $N=\sum_{|p|\leq p_*} (2N_p+1)$ and $M=\sum_{|p|\leq p_*}(2M_p+1)$,
is block-diagonal
with each block  (indexed by $p$) in the form of random partial Fourier matrix, representing the sensing matrix on the dyadic scale $2^p$. 

\commentout{
The sampling scheme \eqref{49'}, \eqref{51} and \eqref{50'} give rise to 
the sampling  range 
\[
x_k\in {z_0\over \om} \pi \Om [-1,1]\supseteq   {z_0\over \om} 2\pi 2^{p_*} [-1,1].
\]
and hence the resolution limit
\beq
\label{56}
2^{-p_*} \geq {2z_0\over A\om}
\eeq
where 
\[
A\equiv  {z_0\over \om} \Om 
\]
is related to 
 is the aperture. The resolution limit \eqref{56} is consistent with
 the classical Abbe or Rayleigh limit, implying no super resolution effect.
 } 
 

 \section{Near-field diffraction with Fourier basis} \label{sec6}
 
Consider near-field  diffraction by a periodic, extended object (e.g. diffraction grating) where the evanescent modes as well as the propagation modes are taken into account. 
Since we can not apply the paraxial approximation, we resort to the Lippmann-Schwinger  equation \eqref{exact'}. 
 
Suppose the  masked object  function is sparse in the 
 the Fourier basis 
 \beq
 \label{0.1}
 V(x)=\sum_{j=-\infty}^{\infty} \hat V_{j}
 e^{i2\pi jx/L}
 \eeq
 where $L$ is the period and only $s$ modes have nonzero amplitudes. 
Suppose that  $\hat V_j=0$ for $j\neq 1,...,N$. 
 
The 2D Green function can be expressed by  the Sommerfeld integral formula 
  \beq
  \label{somm}
 G(\br)={i\over 4\pi}
  \int e^{i\om(|z|\beta(\alpha)+x\alpha)} {d\alpha\over \beta(\alpha)},  \quad \br=(z,x)
  \eeq
  where 
  \beq
  \label{16}
  \beta(\alpha)=\lt\{\begin{array}{ll}
  \sqrt{1-\alpha^2}, & |\alpha|<1\\
  i\sqrt{\alpha^2-1},& |\alpha|> 1
  \end{array}
  \rt.
  \eeq
 (Born and Wolf 1999).  The integrand in (\ref{somm}) with  
  real-valued $\beta$ (i.e. $|\alpha|<1$)
corresponds to the homogeneous wave
and that with imaginary-valued $\beta$ (i.e. $|\alpha|>1$) corresponds
to the evanescent (inhomogeneous) wave which has
an exponential-decay factor $e^{-\om |z| \sqrt{\alpha^2-1}}$. 
Likewise the 3D Green function can be represented by
the Weyl integral formula  (Born and Wolf 1999).  

The signal arriving at the sensor located at $(0, x)$ is given by
the Lippmann-Schwinger equation with (\ref{somm})
\beq
\label{sense}
\int G(z_0, x-x') V(x') dx'&=&
{i\over 2\om}\sum_{j} {\hat V_j\over \beta_j}  e^{i\om z_0\beta_j } e^{i\om \alpha_jx}
\eeq
 where 
\beq
\label{17}
\alpha_j= {2\pi j\over L\om}, & & \beta_j = \beta(\alpha_j).
  \eeq
The subwavelength structure  is encoded in
$\hat V_j$  with $\alpha_j>1$ corresponding to  
  the evanescent modes. 

Let $(0, x_k), x_k=\xi_kL, k=1,...,M$ be the coordinates
of the sampling points where $\xi_k\in [-1/2,1/2]$. In other
words, $L$ is also the aperture (i.e. the sampling range for $x_k$).  
To set the problem in the framework of compressed sensing 
 we 
   set the vector
 $\mbf=(f_j)\in\IC^N$ as 
 \beq
 f_j={i e^{i\om z_0\beta_j }\over 2\om \beta_j}   
 {\hat V_j}.\label{59}
 \eeq
  To avoid a vanishing denominator in (\ref{59}), we assume  that $\alpha_j\neq 1$ and hence $\beta_j\neq 0, \forall j\in \IZ$. 
  This is the case, for instance, when $L\om/(2\pi)$ is irrational.

This gives rise to 
the sensing matrix  $\bPhi$ 
with the entries
\beq
\label{0.6}
\Phi_{kj}&=&e^{i\om\alpha_j x_k}=e^{i2\pi j \xi_k},\quad
k=1,...,M,\quad j=1,...,N
\eeq
\commentout{
 \beq
 \label{1.10}
 A_{jk}&=&\lt\{\begin{array}{ll}
 {e^{i\om z_0 \beta_k}\over \sqrt{n}} e^{i\om \alpha_k \xi_j} ,
 &\mbox{for} \quad \alpha_k\leq 1\\
 {-i\over \sqrt{n}} e^{i\om \alpha_k \xi_j} ,
 &\mbox{for}\quad  \alpha_k > 1
 \end{array}\rt.
 \eeq
 for $
\forall j=1,...,n$.
}
which again is  
the random partial Fourier matrix. 

 A source of instability lurks in the expression \eqref{59} where $\beta_j$ may be complex-valued, corresponding to the evanescent modes. Stability in  inverting the relationship (\ref{59}) requires limiting
 the number of the evanescent modes involved in \eqref{59}. Here the transition is not clear-cut, however. 
 For example, if we demand that 
\beq
|e^{i\om z_0\beta_j}|\geq e^{-2\pi}
\eeq
as the criterion for stable modes, then  the {stable} modes  include  $|\alpha_j|\leq 1$ as well as  $|\alpha_j|>1$ such that
\beq
\label{51}
\om |\beta_j| z_0\leq 2\pi 
\eeq
or equivalently
\beq
\label{resolve}
{|j|\over L}\leq \sqrt{{\om^2\over 4\pi^2}+ {1\over z_0^{2}}}
\eeq
In other words, the number of {\em stably} resolvable modes is proportional to
the probe frequency and inversely proportional to the 
 the distance $z_0$ between the sensor array and
the object. As $z_0$ drops below the wavelength, the subwavelength
Fourier modes of the object can be stably recovered.  
This is the idea behind
 the near-field imaging systems such as  the scanning 
microscopy. 

 \section{Inverse scattering}\label{sec7}

     In the inverse scattering theory, the scattering amplitude is the observable data and 
the main objective  then is to reconstruct
 $\nu$ from the knowledge
of the scattering amplitude.   

\subsection{Pixel basis}\label{sec7.1}\label{sec:is}
To obtain a sensing matrix with compressive sensing properties,  we first make the Born approximation in \eqref{sa} and neglect the scattered field
$u^{\rm s}$ on the right hand side of \eqref{sa}. Our purpose here is
to demonstrate how to coordinate the incidence direction 
and the sampling direction and  create a favorable sensing matrix.

 Consider the  incidence field
 \beq
 \label{inc}
 u^{\rm i}(\br)=e^{i\om\br\cdot \hat\bd}
 \eeq
 where $\hat\bd$ is the incident direction. 
Under 
the Born approximation, we have from \eqref{sa} that 
\beq
\label{born}
A(\hat\br,\hat\bd)=A(\bs)={\om^2\over 4\pi}
\int_{\IR^d}  \nu(\br') e^{-i\om\br'\cdot\bs}d\br'
\eeq
where $ \bs=\hat\br-\hat\bd$ is the scattering vector.
 
We proceed to discretize the continuous system \eqref{born} as before.
Consider  the discrete approximation of
the extended  object $\nu$   
\beq
\label{533}
\nu_{\ell}(\br)=\sum_{\bq\in \cI}
b({\br\over \ell}-\bq ) \nu(\ell\bq)
\eeq
where 
 \beq
 \label{sq}
 b(\br)=\left\{\begin{matrix}
 1,& \br\in [-{1\over 2}, {1\over 2}]^2\\
 0,& \hbox{else}.
 \end{matrix}\right.
 \eeq
is the pixel basis.

\commentout{
where 
 \beq
 \label{sq}
 b(\br)=\left\{\begin{matrix}
 1,& \br\in [-{1\over 2}, {1\over 2}]^2\\
 0,& \hbox{else}.
 \end{matrix}\right.
 \eeq
 }
 
 Define the target vector $\mbf=(f_j)\in \IC^N$ with $f_j=\nu(\ell\bp), \bp=(p_1,p_2) \in \cI,  j=(p_1-1)\sqrt{N}+p_2$. Let $\om_l$ and $ \hat \bd_l$ be the probe frequencies
and directions, respectively, and let $\hat\br_l$ be the
sampling directions for $l=1,...,M$. Let $\bg$ be the data vector with
\[
g_l={4\pi A({ \hat\br_l}-\hat\bd_l) \over \om^2   \hat b({\ell\om_l\over 2\pi}(\hat\br_l-\hat\bd_l))}.
\]
Then the  sensing matrix takes the form
\beq
\label{535}
{ \Phi_{lj}}&=& e^{i\om_l \ell \bq\cdot (\hat\bd_l-\hat\br_l)},\quad \bq=(q_1,q_2)\in \cI, \quad j=(q_1-1)\sqrt{N}+ q_2.
\eeq

\subsection{Sampling schemes}\label{sec7.2}
\label{sec:samp}
Our strategy is to  construct a sensing matrix analogous to the  random partial Fourier matrix. To this end, 
we write the $(l,j)$-entry of the sensing matrix in the form
\beq
e^{i \pi (j_1\xi_l+j_2\zeta_l)},\quad j=(j_1-1)\sqrt{N}+j_2,\quad j_1, j_2=1,...,\sqrt{N},\quad l=1,...,M\nn
\eeq
where $\xi_l, \zeta_l$ are independently and
uniformly distributed in $[-1,1]$.  
Write $(\xi_l,\zeta_l)$ in the polar coordinates $\rho_l, \phi_l$ as 
\beq
(\xi_l,\zeta_l)=\rho_l (\cos\phi_l,\sin\phi_l),\quad 
\rho_l=\sqrt{\xi_l^2+\zeta_l^2} \leq\sqrt{2}
\eeq
and set 
\beqn
\om_l(\cos\theta_l-\cos\tilde\theta_l)& =&\sqrt{2}\rho_l\Om\cos\phi_l\\
\om_l(\sin\theta_l-\sin\tilde\theta_l)&=&\sqrt{2}{\rho_l\Om } 
\sin \phi_l
\eeqn
where $\Om$ is a parameter to be determined later (\ref{75}). 
Equivalently we have
\beq
\label{200}
-\sqrt{2}\om_l\sin{\theta_l-\tilde\theta_l\over 2}\sin{\theta_l+\tilde\theta_l\over 2}& =& \Omega\rho_l\cos\phi_l\\
\sqrt{2} \om_l\sin{\theta_l-\tilde\theta_l\over 2}\cos{\theta_l+\tilde\theta_l\over 2}&=&\Omega\rho_l\sin \phi_l. 
\label{201}
\eeq
This set of equations determines the single-input-$(\theta_l,\om_l)$-single-output-$\tilde\theta_l$ 
mode of sampling.

The following   implementation of  (\ref{200})-(\ref{201}) is natural.  Let the sampling angle $\tilde\theta_l$ be related
to the incident  angle $\theta_l$ via 
\beq
\label{ch1}
\theta_l+\tilde\theta_l=2\phi_l+\pi,
\eeq
and set the frequency $\om_l$ to be
\beq
\label{ch2}
\om_l= {{  \Omega \rho_l}\over \sqrt{2} \sin{\theta_l-\tilde\theta_l\over 2}}.
\eeq
Then the entries {  (\ref{535})} of the sensing matrix $\bA$ 
have the form
\beq
e^{i\sqrt{2}\Om \ell(j_1\xi_l+j_2\zeta_l)}, \quad
l=1,...,n,\quad j_1, j_2=1,...,\sqrt{N}.\label{17-2}
\eeq
By the square-symmetry of the problem,
it is clear that the relation (\ref{ch1}) can be generalized to 
\beq
\label{ch1'}
\theta_l+\tilde\theta_l=2\phi_l+\eta \pi,\quad \eta\in \IZ.   
\eeq
On the other hand, the symmetry of the square lattice 
should not  play a significant role and
hence we expect the result to be 
insensitive to 
any {\em fixed} $\eta\in \IR$, independent of $l$, as
long as (\ref{ch2}) holds.
Indeed this is confirmed by numerical simulations.

Let us focus on  two specific measurement schemes.

\noindent{\em Backward sampling.} 
This scheme employs  $\Om-$band limited probes, i.e.
$\om_l\in  [-\Om,\Om]$. 
This and (\ref{ch2}) lead to
the constraint:
\beq
\label{const}
{\lt| \sin{\theta_l-\tilde\theta_l\over 2}\rt|}\geq {\rho_l\over\sqrt{2}}. 
\eeq

The simplest way  to satisfy (\ref{ch1}) and (\ref{const}) is
to set 
\beq
\label{41}\phi_l&=&\tilde\theta_l=\theta_l+\pi,\\
\label{42} \om_l&=& {{  \Om\rho_l}\over \sqrt{2}}
\eeq
$l=1,...,n$. In this case the scattering amplitude is always sampled in 
the back-scattering direction. This resembles
the synthetic aperture imaging which has been 
previously analyzed under the paraxial approximation
in Fannjiang {\em et al.} 2010. 
 In contrast, the forward
scattering direction with $\tilde\theta_l=\theta_l$ almost surely violates
the constraint  (\ref{const}). 

\noindent {\em Forward sampling.}
 This scheme employs  single  frequency probes no less
 than $\Omega$: 
\beq
\label{20}
\om_l=\gamma\Omega,\quad \gamma\geq 1,\quad   l=1,...,n.\eeq
To satisfy (\ref{ch1'}) and (\ref{ch2}) we set 
\commentout{
and set
\[
\sin{\theta_l-\tilde\theta_l\over 2}={\rho_l\over \gamma \sqrt{2}}
\]
which is solved by
}
\beq
\label{21}
\theta_l=\phi_l+{\eta\pi\over 2} +\arcsin{\rho_l\over \gamma\sqrt{2}}\\
\tilde\theta_l=\phi_l+{\eta\pi\over 2}-\arcsin{\rho_l\over \gamma\sqrt{2}}\label{22}
\eeq
with $\eta\in \IZ$. 
The difference between the incident angle and
the sampling angle is 
\beq
\label{25}
\theta_l-\tilde\theta_l=2\arcsin{\rho_l\over \gamma\sqrt{2}}
\eeq
which diminishes as $\gamma\to\infty$. In other words, in the high frequency limit, the sampling angle approaches the
incident  angle. This resembles the setting of  the X-ray tomography.

\commentout{
\bigskip

\noindent {\bf Scheme III.}
This scheme employs probes of unlimited frequency band.
Let $\theta_l$ be $n$ arbitrary distinct numbers in $ [-\pi,\pi]$ 
and let $\tilde\theta_l$ and $\om_l$ be determined
by (\ref{ch1'}) and (\ref{ch2}), respectively. 
The possibility of having a small divisor in (\ref{ch2})
renders  the bandwidth  unlimited in principle. 
}

In summary, let $\xi_l, \zeta_l$ be independently and
uniformly distributed in $[-1,1]$ and 
let $(\rho_l,\phi_l)$ be the polar coordinates
of  $(\xi_l,\zeta_l)$, i.e. 
\[
(\xi_l,\zeta_l)=\rho_l(\cos\phi_l,\sin\phi_l).
\]
Then with with \beq
\label{75}
\Omega\ell=\pi/\sqrt{2}
\eeq
both forward and backward samplings give rise to the random partial Fourier sensing matrix.
\subsection{Coherence bounds for single frequency}\label{sec:is2}\label{sec7.3}

As in Section \ref{sec:blo} we let
the point scatterers be continuously distributed over a finite domain, not necessarily
on a grid. Any computational imaging would involve some underlying, however refined, grid. Hence  let us assume that there is an underlying, possibly highly refined and unresolved, grid of spacing $\ell\ll \om^{-1}$ (the reciprocal of probe frequency).  

We shall focus on the monochromatic case with $\om_l = \om, l = 1,...,M$.

Recall the sensing matrix continues of the form \eqref{535} which now becomes
\beq
\label{535''}
\phi_{lj}&=& e^{i\om\ell \bp\cdot (\hat\bd_l-\hat\br_l)},\quad j=(p_1-1)\sqrt{N}+ p_2,\quad \bp\in \cI.
\eeq
In other words, the measurement diversity comes entirely from the variations of 
the incidence and detection directions.
We assume that the $n$ incident directions  and the $m$ detection directions are each independently chosen according to some distributions with the total number of data $M=nm$ fixed.  

 \begin{theorem}\label{thm4}
 {\bf (2D case)}. 
 Suppose the incident and sampling angles are randomly, independently and
 identically distributed according to the probability density functions $\pdfi(\theta)\in C^1$ and $\pdfs(\theta)\in C^1$, respectively. 
Suppose
\beq
\label{m-2}
N\leq {\ep\over 8} e^{K^2/2},\quad \ep, K>0.
\eeq

Set $L=\ell |\bp-\bq|$ for any $ \bp, \bq\in \cI$. Then   the sensing matrix satisfies the pairwise coherence bound
\beq
\label{mut}
\mu_{\bp,\bq}< \lt(
\bar\mu^{\rm i}+{\sqrt{2} K\over \sqrt{n}}\rt)
\lt(\bar\mu^{\rm s}+{\sqrt{2}K\over \sqrt{m}}\rt)
\eeq
 with probability greater than $(1-\ep)^2$
 where 
 \beq
 \label{21-3}
&\bar\mu^{\rm i}\leq {c}{ {(1+\om L)}^{-1/2}} \sup_\theta\left\{|\pdfi(\theta)|, \left|{d\over d\theta} \pdfi(\theta)\right|\right\},&\\
&\bar\mu^{\rm s}\leq {c}{(1+\om L)^{-1/2}}\sup_\theta\left\{|\pdfs(\theta)|, \left|{d\over d\theta} \pdfs(\theta)\right|\right\},&\label{21-4}
 \eeq
{  with a positive constant $c$}.
\label{thm3-1}
\end{theorem}

In 3D, the coherence bound can be improved with a faster decay rate in terms of
$\om L\gg 1$ as stated below.

 \begin{theorem}\label{thm6}
{\bf (3D case)}.  Assume (\ref{m-2}).
 Suppose the incidence and sampling directions, parametrized by the polar  angle
 $\theta\in [0,\pi]$ and the azimuthal angle $ \phi\in [0,2\pi]$,  are randomly,  independently and identically
 distributed. Let $\pdfi(\theta)\in C^1$ and $\pdfs(\theta)\in C^1$ be the marginal density functions of the incident and sampling polar angles,  respectively.

Let $L=\ell|\bp-\bq|$. Then   the sensing matrix  satisfies the pairwise coherence bound
\beq
\label{mut'}
\mu_{\bp,\bq} < \lt(
\bar\mu^{\rm i}+{\sqrt{2} K\over \sqrt{n}}\rt)
\lt(\bar\mu^{\rm s}+{\sqrt{2}K\over \sqrt{m}}\rt)
\eeq
 with probability greater than $(1-\ep)^2$
 where 
 \beq
 \label{21-3'}
&\bar\mu^{\rm i}\leq c (1+\om L)^{-1} \sup_\theta\left\{|\pdfi(\theta)|, \left|{d\over d\theta} \pdfi(\theta)\right|\right\}&\\
& \bar\mu^{\rm s}\leq {c   ( 1+\om L)^{-1}
\sup_\theta\left\{|\pdfs(\theta)|, \left|{d\over d\theta} \pdfs(\theta)\right|\right\}}.  &
 \eeq
\end{theorem}\begin{remark}
The original statements of the theorems (Fannjiang 2010b, Theorems 1 and 6) 
have been adapted to the present context of off-grid objects. The original proofs, however, carry over here {\em verbatim} upon minor change of notation. 
\end{remark}

\begin{remark}\label{rmk6}
When the sampling directions are randomized and
the incidence directions are deterministic, then the coherence bounds  {  \eqref{mut}} and \eqref{mut'} hold with  the first factor
on the right hand side removed.
\end{remark}
According to Remark \ref{rmk6}, we have the pairwise coherence bound:
\beq
\hbox{(2D)}& \mu_{\bp,\bq}&\leq {c}{(1+\om L)^{-1/2}}\sup_\theta\left\{|\pdfs(\theta)|, \left|{d\over d\theta} \pdfs(\theta)\right|\right\}+{\sqrt{2}K\over \sqrt{M}}\label{110}\\
\hbox{(3D)}& \mu_{\bp,\bq}&\leq {c}{(1+\om L)^{-1}}\sup_\theta\left\{|\pdfs(\theta)|, \left|{d\over d\theta} \pdfs(\theta)\right|\right\}+{\sqrt{2}K\over \sqrt{M}}  \label{111}
\eeq
which is an estimate  of the coherence pattern of the sensing matrix.
Hence, if $L$ is unresolvable (i.e. $\om L\leq 1 $), 
 the corresponding pairwise coherence parameter is high and
when if $L$ is well-resolved  (i.e. $\om L\gg 1$)  the corresponding pairwise coherence parameter is low. A typical  coherence band has  a coherence radius $\cO(\om^{-1})$ according to
\eqref{110}-\eqref{111}. 

\begin{figure}
\centering
\includegraphics[width=6cm,height=5cm]{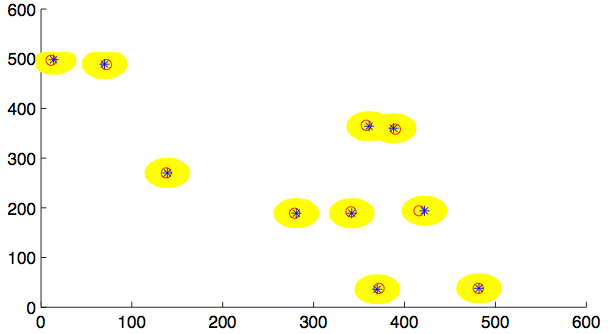}\hspace{1cm}
\includegraphics[width=6cm,height=5cm]{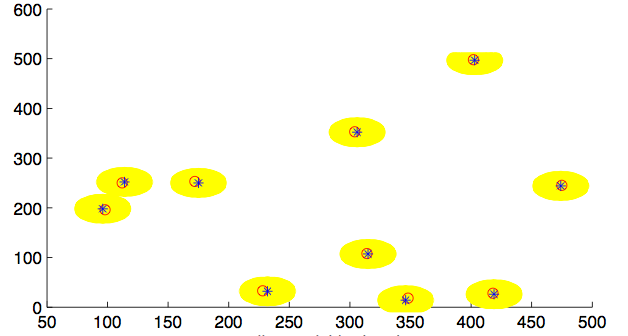}
\caption{Two instances of BOMP reconstruction: red circles are the exact locations, blue asterisks are recovered locations and the yellow
patches are the coherence bands around the objects.}
\label{fig-2d}
\end{figure}

Therefore, if the point objects are well separated in the sense that
any pair of objects are larger than $\om^{-1}$ then the same BLO- and BLOT-based techniques discussed in Section \ref{sec:blo} can be used to
recover the {\em masked}  object support and amplitudes. For a simple illustration,  Figure \ref{fig-2d} shows two instances of reconstruction by BOMP. 
The recovered objects (blue asterisks) are close to the true objects (red circles)
well within the coherence bands (yellow patches).

\section{Inverse multiple scattering}\label{sec8}
 
In this section, we  present an approach to compressive imaging of multiply scattering point scatterers.
First consider the multiple scattering effect with just a single illumination, i.e. $n=1$ and $M=m$. 

Note that the original object support is the same as the masked object support.
With the support accurately recovered, let us consider how to unmask the objects
and recover the true objects. 

Define the incidence and full field vectors
at the locations of the objects:
\beqn
 \bu^{\rm i}&=&(u^{\rm i}(\br_{1}),...,u^{\rm i}(\br_{s}))^T\in \IC^s\\
\bu&=&(u(\br_{1}),...,u(\br_{s}))^T\in \IC^s.
\eeqn

Let $\mathbf{\Gamma}$ be the $s\times s$ matrix 
\[
\mathbf{\Gamma}=[(1-\delta_{jl})G(\br_{j},\br_{l})]
\]
 and $\cV$ the diagonal matrix 
 \[
 \cV={\rm diag}(\nu_{1},...,\nu_{s}). 
 \]
The full field is determined by the Foldy-Lax equation (Mishchenko {\em et al.} 2006) 
\beq
\label{fl2}
\bu=\bu^{\rm i}+\om^2 \mathbf{\Gamma}\cV \bu
\eeq
from which  we obtain the full field
\beq
\bu&=&
\lt(\bI-\om^2\mathbf{\Gamma}\cV\rt)^{-1}
\bu^{\rm i} \label{ill}
\eeq
and the masked objects 
\beq
\label{ls3}
\mbf=\cV \bu&=&\cV\lt(\bI-\om^2 \mathbf{\Gamma}\cV\rt)^{-1}\bu^{\rm i}\\
&=&\lt(\bI- \om^2\cV\mathbf{\Gamma}\rt)^{-1}\cV \bu^{\rm i}\nn
\eeq 
provided that $\om^{-2}$ is not an eigenvalue of 
$\mathbf{\Gamma}\cV$. 

Hence by \eqref{ls3} we have
\beq
 \label{76'}
\lt(\bI-\om^2 \cV\mathbf{\Gamma}\rt)  \mbf=\cV \bu^{\rm i}.
 \eeq
The true objects $\nu$ can then be recovered by solving \eqref{76'} as
 \beq
\label{exact}
\nu={ \mbf\over \om^2 \mathbf{\Gamma} \mbf+\bu^{\rm i}}
\eeq
where the division  is carried out entry-wise (Hadamard product).
\subsection{Joint sparsity}\label{sec8.1}

With the total number of data $M=nm$ fixed the coherence bounds \eqref{mut} and \eqref{mut'} is optimized with $n\sim m\sim \sqrt{M}$. To take advantage of this result, we should  deploy multiple incidence fields for which 
the formula \eqref{exact} is no longer valid. 

Multiple illuminations give rise to multiple data vectors $\bg_j$  and multiple masked object vectors $\mbf_j, j=1,...,n$ each of which is masked by a unknown field $\bu_j$.
However, all  masked object vectors give rise to  the same sensing matrix  
\[\Phi_{lj}=e^{-i\om\ell \bp\cdot \hat\br_l},\quad j=(p_1-1)\sqrt{N}+ p_2,\quad \bp\in \cI.
\]

Since every masked object vector shares the same support as the true object vector, 
this is a suitable setting for the application of joint sparsity techniques discussed 
in Sections \ref{sec3.2} and \ref{sec3.3}.
 
Compiling  the masked object vectors as $\bF=[\mbf_1,..., \mbf_n]\in  \IC^{m\times n}$ and the data vectors as $\bG=[\bg_1,...,\bg_n]\in   \IC^{m\times n}$, we obtain the imaging equations
\beq
\bG=\bA\bF+\bE \label{117}
\eeq
where $\bE$ accounts for noise. 
When the true objects are widely separated, we have two ways to proceed as follows.\\

\noindent{\bf 1) BPDN-BLOT for joint sparsity.} In the first approach, 
we  use BPDN for joint sparsity \eqref{jbp} 
 with $\bPhi_j=\bPhi, { \forall j}, \cL=0$ to solve the imaging equation \eqref{117}.
 Let $ \bF_*=( \mbf_{1*},...,\mbf_{n*})$ be the solution. We then apply the BLOT technique (Algorithm 5) to improve $\bF_*$. In order to enforce the joint sparsity structure, we modify Algorithm 5 as follows.
 
 First, we modify the LO algorithm to account for joint sparsity.
 \begin{center}
   \begin{tabular}{l}
   \hline  
   \centerline{{\bf Algorithm 7.}\quad LO for joint sparsity}  \\ \hline
    Input: $\bA_1,...,\bA_n,\bG, \eta>0,  S^0=\{i_1,\ldots,i_s\}$.\\
Iteration:  For $k=1,2,...,s$.\\
  \quad  1) $\bF^k = \hbox{arg} \min\|
     [\bA_1\bh_1,...,\bA_n\bh_n]-\bG\|_{\rm F}$ s.t. 
$ \cup_j\hbox{supp}(\bh_j){ \subseteq }(S^{k-1} \backslash \{i_k\})\cup \{i'_k\}, $  $  i'_k\in B_\eta(\{i_k\})$.
\\
 \quad 2) $S^k=\hbox{supp}(\bF^k)$.\\
    Output:  $S^s$.\\
    \hline
   \end{tabular}
\end{center}

\bigskip

Next, we modify the BLOT algorithm to account for joint sparsity.
 \begin{center}
   \begin{tabular}{l}
   \hline
   \centerline{{\bf Algorithm 8.}\quad BLOT for joint sparsity}  \\ \hline
    Input: $\mbf_1,...,\mbf_n$, $\bA_1,...,\bA_n, \bG, \eta>0$.\\
    Initialization:  $S^0=\emptyset$.\\
Iteration:  For $k=1,2,...,s$.\\
\quad 1) $i_k= \hbox{arg}\,\,\max_j \|\mbf_j\|_2, k\not\in B^{(2)}_\eta(S^{k-1}) $.\\
 \quad 2)  $S^k=S^{k-1}\cup\{i_k\}$.\\
    Output: $\bF_*=\hbox{arg}\min \| [\bA_1\bh_1,...,\bA_n\bh_n]-\bG\|_{\rm F}$, $\cup_j\hbox{supp}(\bh_j){\subseteq} \hbox{JLO}(S^s)$ {  where $\hbox{JLO} (S^s)$}\\
    \quad \quad  \quad \quad  {  is the output of Algorithm 7 with the $s$-th iterate $S^s$ of BLOT as input.}\\
     \hline
   \end{tabular}
\end{center}

\bigskip

\noindent{\bf 2) BLOOMP for joint sparsity.}  In the second approach, we propose the following  joint sparsity version of 
BLOOMP. 

\begin{center}
   \begin{tabular}{l}
   \hline
   \centerline{{\bf Algorithm 9.} BLOOMP for joint sparsity} \\ \hline
   Input: $\bA_1,...,\bA_n, \bG,\eta>0$\\
 Initialization:  $\bF^0 = 0, \bR^0 = \bG$ and $S^0=\emptyset$ \\ 
Iteration: For  $k=1,...,s$\\
\quad 1) $i_{\rm max} = \hbox{arg}\max_{i}\sum^J_{j=1}|\Phi^\dagger_{j,i}\br^{k-1}_j |, i \notin B^{(2)}_\eta(S^{k-1}) $, \hbox{where $\Phi^\dagger_{j,i}=$ conjugate transpose  of $\hbox{\rm col}_i(\bPhi_j)$}. \\
  \quad      2) $S^{k} = \hbox{JLO}(S^{k-1} \cup \{i_{\rm max}\})$ where $\hbox{JLO}$ is the output of Algorithm 7.\\
  \quad  3) $[\mbf^k_1,...,\mbf^k_n]= \hbox{arg}  \min_\bH \|
   [\bA_1\bh_1,...,\bA_n\bh_n]-   \bG\|_{\rm F}$ s.t. $\cup_j\hbox{supp}(\bh_j$) ${  \subseteq } S^k$ \\
  \quad   4) $[\br^k_1,...,\br^k_n] = \bG-  [{ \bA_1}\mbf^k_1,...,{  \bA_n}\mbf^k_n]$\\
 Output: $\bF_*=[\mbf^s_1,...,\mbf^s_n]$. \\
 \hline
   \end{tabular}
\end{center}


After the first stage of either approach, we obtain an estimate of the object support as well as the amplitudes of masked objects. In the second stage, we estimate the
true object amplitudes. If we use the formula \eqref{exact} for each incident wave $\bu^{\rm i}_j$, we  end up with $n$ amplitude estimates \beq
\nn
{ \mbf_{j*}\over \om^2 \mathbf{\Gamma}\mbf_{j*}+\bu^{\rm i}_j},\quad j=1,...,n
\eeq 
 that are typically inconsistent.
Least squares is the natural way to solve this over-determined system 
{  and obtain the object estimate }
\[
\nu_*=\hbox{\rm arg}\min_{\mathbf{v}}\sum_{j=1}^n\| (\om^2 \mathbf{\Gamma}\mbf_{j*}+\bu^{\rm i}_j)\mathbf{v}-\mbf_{j*}\|_2^2. 
\]

\section{Inverse Scattering with Zernike basis}\label{sec9}
In this section, we discuss a basis for representing extended objects in
the scattering geometry and its application to compressive inverse scattering.
We shall make the Born approximation. 

A well known orthogonal basis for representing 
an extended object with a compactly support (e.g. the unit disk) 
is the product of Zernike polynomials $R^m_n$  and trigonometric functions 
\beq
V^m_n(x,y)=V^m_n(\rho\cos\theta,\rho\sin\theta)=R^m_n(\rho)e^{im\theta},\quad x^2+y^2\leq 1
\eeq
where $m\in \IZ, n\in \IN$, $n\geq |m|$  and $n-|m|$ is even. 
We refer to $V^m_n$ as the { Zernike functions} of order $(m,n)$ (Born and Wolf 1999). 
These Zernike functions are very useful in optics because the lowest few terms of a Zernike expansion have a simple optical interpretation (Dai and {  Mahajan 2008}). In addition, a
Zernike expansion usually has a superior rate of convergence (hence sparser) compared with other
expansions such as a Bessel-Fourier or Chebyshev-Fourier expansion (Boyd and Yu 2011 and Boyd and Petschek 2014). 

We show now that the Zernike basis also results in a better coherence parameter (hence better resolution) than the pixel basis. 
The Zernike polynomials are given explicitly by the formula
\beq
R^{m}_n(\rho)&=& {1\over ({n-|m|\over 2})\! \rho^{|m|}}
\left[{d\over d(\rho^2)}\right]^{n-|m|\over 2}
\left[(\rho^2)^{n+|m|\over 2}(\rho^2-1)^{n-|m|\over 2}\right] 
\eeq
which are $n$-th degree  polynimials in $\rho$ and  normalized such that $R^m_n(1)=1$ for all permissible values of $m,n$. 
The Zernike polynomials satisfy the following properties
\beq
\int^1_0R^m_n(\rho)R^m_{n'}(\rho)\rho d\rho&=& {\delta_{nn'}\over 2(n+1)} \label{46}\\
\int^1_0 R^m_n(\rho) J_m(u\rho) \rho d\rho&=& (-1)^{n-m\over 2}
{J_{n+1}(u)\over u}\label{47}
\eeq
where $J_{n+1}$ is the $(n+1)$-order Bessel function of the first kind.
As a consequence of (\ref{46}), the Zernike functions satisfy the orthogonality property
\beq
\int_{x^2+y^2\leq 1} \overline{V^m_n(x,y)} V^{m'}_{n'}(x,y)dxdy
={{  \pi}\over n+1}\delta_{mm'}\delta_{nn'}.
\eeq

Writing $\bs=s(\cos\phi,\sin\phi)$, let us compute the matrix element for the scattering amplitude \eqref{born} as follows.
\beq
\label{49}\int_{x^2+y^2\leq 1} \overline{V^m_n(x,y)}e^{-i\om{ \bs}\cdot(x,y)}
dxdy&=&\int^1_0\int_0^{2\pi}e^{i\om s \rho \cos{(\phi+\theta)}}{  R^m_n(\rho)e^{-im\theta}}d\theta \rho d\rho\\
&=&\int^1_0\int^{2\pi}_0  e^{i\om s \rho \cos\theta} e^{-im\theta} d\theta R^m_n(\rho) \rho d\rho e^{im \phi} \nn\\
&=&2 \pi i^n e^{im\phi}\int^1_0 J_m(\om s \rho) R^m_n(\rho) \rho d\rho \nn
\eeq
by the definition of Bessel function 
\[
J_m(z)={1\over \pi i^m} \int^\pi_0 e^{iz\cos\theta} \cos{(m\theta)}d\theta.
\]
Using the property (\ref{47}), we then obtain from \eqref{49} that 
\beq
\label{50}\int_{x^2+y^2\leq 1} \overline{V^m_n(x,y)}e^{-i\om{  \bs}\cdot(x,y)}
dxdy&=&2\pi i^m  (-1)^{n-m\over 2}e^{im\phi} 
{J_{n+1}(\om s)\over \om s}
\eeq
which are the sensing matrix elements with all permissible $m,n$.
Note that the columns of the sensing matrix are indexed by 
the permissible $m\in \IZ, n\in \IN$ with the constraint that  $n\geq |m|$  and $n-|m|$ is even.

Let  the scattering vector $\bs=\hat\br-\hat\bd$  be parametrized  as 
 \[
 \bs_{jk}=s_j(\cos\phi_k,\sin\phi_k), \quad  j,k=1,...,\sqrt{M} 
 \]
such that $\{\phi_k\}$ are independently and identically  distributed  uniform random variables  on  $[0,2\pi]$ and $\{s_j\}$ 
are independently distributed on $[0,2]$ according to
the linear density function  $f(r)=r/2$.  As a result, $z_j=\om s_j$ are independently and identically distributed
on $[0,2\om]$ according to a linear density function.

Calculation of the coherence parameter between the columns corresponding to $(m,n)\neq (m',n')$ gives the following expression 
\beq
\left({1\over \sqrt{M}}\sum_{j=1}^{\sqrt{M}} {J_{n+1}(\om s_j)\over \om s_j} 
 {J_{n'+1}(\om s_j)\over \om s_j} \right)\left({1\over \sqrt{M}}
 \sum^{\sqrt{M}}_{  k=1} e^{i(m-m')\phi_k}\right). \nn
 \eeq
 
Recall that for $p,q\in \IN$ 
 \beq
 \int^\infty_0J_{p}(z)J_{q}(z){dz\over z}=
 \left\{\begin{matrix} 
 0, & p\neq q\\
 {1\over 2p},& p=q
 \end{matrix} \right. \label{53}
\eeq
(Abramowitz and Stegun 1972, formula 11.4.6). 
 For $M\gg 1$, we have by the law of large numbers
  \beq
\label{52} {1\over \sqrt{M}}\sum_{j=1}^{\sqrt{M}} {J_{n+1}(\om s_j)\over \om s_j} 
 {J_{n'+1}(\om s_j)\over \om s_j}&\sim& \IE\left[{J_{n+1}(\om r)\over \om r} 
 {J_{n'+1}(\om r)\over \om r}\right]\\
 &=&{1\over 2\om^2} \int^{2\om}_0{J_{n+1}(z)} 
 {J_{n'+1}(z)}{dz\over z}\nn
  \eeq
and
\beq
\label{54} {1\over \sqrt{M}}
 \sum^{\sqrt{M}}_{  k=1} e^{i(m-m')  \phi_k}\sim
 \IE e^{i  (m-m')\phi}&=&\int^{2\pi}_0 e^{i(m-m')\phi}g(\phi) d\phi\\
 &=&\delta_{mm'}.\nn
 \commentout{
 &=&\left\{\begin{matrix}
 {1\over i2\pi (m-m')} \left(e^{i2\pi\alpha (m-m')}-1\right),& m\neq m'\\
 \alpha,& m=m'.
 \end{matrix}\right.\nn
 }
 \eeq
When $m\neq m'$, the two columns are orthogonal and the pairwise coherence parameter is zero. 
 When $n\neq n'$, the right hand side of \eqref{52} becomes  $\cO(\om^{-3})$
 in view of \eqref{53} and the fact that  the Bessel functions $J_n(z)$ decay like $z^{-1/2}$ for $z\gg 1$. From \eqref{53} and  \eqref{52} with $n=n'$, we see that the 2-norm of the columns
 is $\cO(\om^{-2})$. 
After dividing \eqref{52} with $n\neq n'$ by the 2-norm of the columns 
 the coherence parameter  scales at worst like $\om^{-1}$ (for 
 $m=m', n\neq n'$).
 
 Notice that  this decay date of the coherence parameter  is faster
 than the $\om^{-1/2}$ behavior in  \eqref{mut}-\eqref{21-3}. Hence, imaging with
 the Zernike basis possess better resolution capability   than 
 with the pixel basis, all else  being equal. 
 
 \section{Interferometry with incoherent sources}\label{sec10}
 In this last section, we discuss the compressive sensing application to
 optical interferometry in astronomy which has a similar mathematical structure
 to that of the inverse scattering \eqref{535''} under the Born approximation. 
 
In astronomy, interferometry often  deals with signals  emitted from incoherent sources.  
In this section, we present compressive sensing approach to such a problem.
With the help of the van Cittert-Zernike theorem, the sensing matrix has a structure not
unlike what we discuss above. 

Suppose the field of view is small enough to be identified
with a planar patch of the celestial sphere $\cP$, called the object plane. Let $I(\bs)$ be the radiation intensity from the point
$\bs$ 
on the object plane $\cP$. Let $n$ antennas be located
in a square of size $L$ on the sensor plane parallel to $\cP$ with locations  $L\br_j, j=1,...,n$
where $\br_j\in [0,1]^2$. 
Then by van Cittert-Zernike theorem (Born and Wolf 1999) 
the measured visibility $v(\br_j-\br_k)$ is given by
the Fourier integral
\beq
v (\br_j-\br_k)=\int_{\cP} I (\bs) e^{i \om\bs\cdot(\br_j-\br_k)L}d\bs.\label{76}
\eeq

Consider   the discrete approximation of
the extended  object  $I$ with the pixel basis on the grid $\ell\cI$ 
\beq
\label{533'}
I_\ell(\br)=\sum_{\bq\in \cI}
b({\br\over \ell}-\bq ) I(\ell\bq)
\eeq
where $b$ is given in \eqref{sq}
 and \beq
\label{latt}
\cI=\{\bp=(p_1, p_2): p_1, p_2= 1,...,\sqrt{N}\}.
\eeq
Substituting \eqref{533'} into \eqref{76} we   obtain  the discrete sum
\beq
v (\br_j-\br_k)= \ell^2\hat b \left({\om\ell L\over 2\pi}(\br_k-\br_j)\right) \sum_{l=1}^NI_le^{i \om \bp\cdot(\br_j-\br_k)\ell L}, \label{2}
\eeq
where  $l, \bp$ are related by $ l=(p_1-1)\sqrt{N}+ p_2$ and
\[
\hat b(\xi,\eta)={\sin{(\pi \xi)}\over \pi \xi} {\sin{(\pi \eta)}\over \pi \eta}. 
\]
For every pair $(j,k)$ of sensors we measure and collect the 
interferometric datum $v(\br_j-\br_k)$ and  we want to determine $I$ from the collection of $n(n-1)$
real-valued data.

Let us rewrite eq. (\ref{2}) in the form \eqref{u1}. In contrast to {  \eqref{pix}}, we set
\beq
\label{6}
\ell={\pi\over \om L}
\eeq
to account for the ``two-way"  structure in the imaging equation \eqref{2}. 
Note that $\ell$ is the resolution length on the celestial sphere and hence dimensionless.

 Let
$\mbf=(f_i)\in \IR^N$ be the unknown object vector, i.e. $f_i=\ell^2 I_i$. 
Let 
$\bg=(g_l)\in \IR^{M}, M=n(n-1)/2,$ 
\beqn
g_l&= &{1\over \hat b \left((\br_k-\br_j)/2\right) }\lt\{ \begin{matrix}
\Re\lt[v(\br_j-\br_k)\rt], & l=(2n-j)(j-1)/2+k,\quad j<k=1,...,n\\
\Im\lt[v(\br_j-\br_k)\rt], & l=n(n-1)/2+(2n-j)(j-1)/2+k,\quad j<k=1,...,n
\end{matrix}
\rt.
\eeqn
be the data vector where $\Re$ and $\Im$ stand for, respectively,  the real and
imaginary parts. 
The sensing matrix $\bA\in \IR^{M\times N}$ now takes the form 
\beq
\label{4}
\Phi_{il}&=&\lt\{\begin{matrix}
 \cos\lt[2\pi \bp_l\cdot(\br_j-\br_k)\rt], & i=(2n-j)(j-1)/2+k,\quad j<k
\\
 \sin\lt[2\pi \bp_l\cdot(\br_j-\br_k)\rt], & i=n(n-1)/2+ (2n-j)(j-1)/2+k,\quad j<k
 \end{matrix}\rt.
\eeq
which  is no longer the simple random partial Fourier matrix for 2D 
as the baselines $\br_j-\br_k$ are related to one another. Nevertheless \eqref{4} has a similar structure to that of the inverse scattering \eqref{535''} when the transmitters and receivers are co-located. Note that
as $(\br_k-\br_j)/2\in [-1/2, 1/2]^2$ the denominator $\hat b \left((\br_k-\br_j)/2\right)$
in the definition of $g_l$ does not vanish. 

Next we give an upper bound for the coherence parameter. 
For the pairwise coherence for columns $i,i'$ corresponding to $\bp,\bp\in \cI$, we have
the following calculation
\beqn
\mu(i,i')&=&{2\over n(n-1)}\Big|\sum_{j<k} \cos\lt[2\pi \bp\cdot(\br_j-\br_k)\rt]
 \cos\lt[2\pi \bp'\cdot(\br_j-\br_k)\rt]\\
 &&\hspace{2cm}+\sin\lt[2\pi\bp\cdot(\br_j-\br_k)\rt] \sin\lt[2\pi\bp'\cdot(\br_j-\br_k)\rt]\Big|\\
 &=&{2\over n(n-1)}\Big|\sum_{j<k} \cos\lt[2\pi (\bp-\bp')\cdot(\br_j-\br_k)\rt]\Big|\nn\\
 &=&{1\over n(n-1)}\Big|\sum_{j\neq k} \cos\lt[2\pi (\bp-\bp')\cdot(\br_j-\br_k)\rt]\Big|
 \eeqn
 First we claim:
 \beqn
 \mu(i,i')&=&{1\over n(n-1)}\Big|\Big|\sum^n_{j=1} e^{i 2\pi (\bp-\bp')\cdot{ \br_j}}\Big|^2-n\Big|.
  \eeqn
 This follows from the calculation 
 \beqn
\Big|\sum^n_{j=1} e^{i 2\pi (\bp-\bp')\cdot{\br_j}}\Big|^2-n&=&{ \sum_{j\neq k} e^{i 2\pi (\bp-\bp')\cdot(\br_j-\br_k)}}\\
&=&\sum_{j\neq k} { \cos}\lt[2\pi (\bp-\bp')\cdot(\br_j-\br_k)\rt]+i
 \sin\lt[2\pi (\bp-\bp')\cdot(\br_j-\br_k)\rt]\\
 &=&\sum_{j\neq k} \cos\lt[2\pi (\bp-\bp')\cdot(\br_j-\br_k)\rt]
\eeqn

\commentout{
\begin{figure}[t!]
\centering 
\subfigure[Coherence parameter versus number of sensors]{
\includegraphics[width=6.5cm]{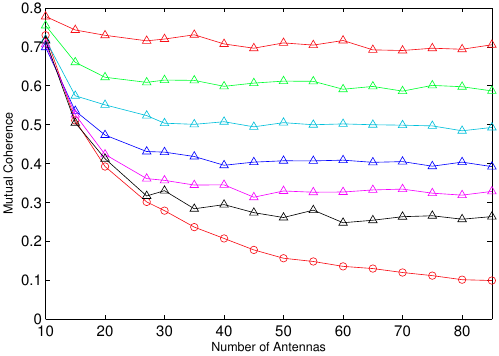}
}
\subfigure[Success rate versus sparsity]{
\includegraphics[width=6.5cm]{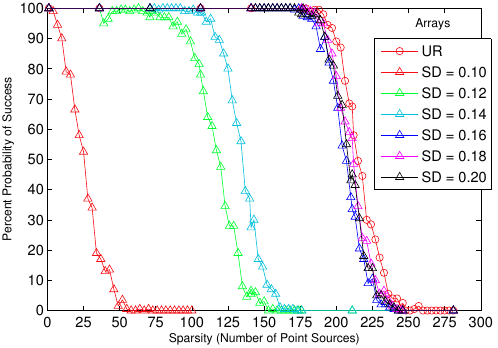}
}
\caption{(C. Fannjiang 2013) (a) Coherence parameter versus the number of sensors; (b) success rate for OMP reconstruction of randomly distributed  point sources of specified sparsity. In the legend, UR stands for uniformly distributed random array of sensors. The rest of curves are for normally distributed array of sensors of specified Standard Deviation (SD).
The number of sensors is 27.}
\label{fig11}
\end{figure}

 \begin{figure}[t]
\centering 
\subfigure[original object (3C75)]{
\includegraphics[width=6cm]{chap2figs/ngc.png}
}\hspace{1cm}
\subfigure[URA]{
\includegraphics[width=6cm]{chap2figs/ngc-ura.png}
}
\vspace{1cm}
\subfigure[NRA (SD $=0.14$)]{
\includegraphics[width=6cm]{chap2figs/ngc-gaussian014.png}
}\hspace{1cm}
\subfigure[NRA (SD $=0.18$)]{
\includegraphics[width=6cm]{chap2figs/ngc-gaussian018.png}
}
\caption{(C. Fannjiang 2013) OMP reconstruction with the pixel basis of (a) the object {\bf 3C75}, using  (b) URA, (c) NRA with SD=0.14 and (d) NRA with SD=0.18. The number of sensors is 100.}
\label{fig12}
\end{figure}
}

\commentout{
 \begin{figure}[t]
\centering 
\subfigure[original object (3C75)]{
\includegraphics[width=6cm]{ngc.pdf}
}\hspace{1cm}
\subfigure[URA]{
\includegraphics[width=6cm]{ngc-ura.pdf}
}
\vspace{1cm}
\subfigure[NRA (SD $=0.14$)]{
\includegraphics[width=6cm]{ngc-gaussian014.pdf}
}\hspace{1cm}
\subfigure[NRA (SD $=0.18$)]{
\includegraphics[width=6cm]{ngc-gaussian018.pdf}
}
\caption{(C. Fannjiang 2013) OMP reconstruction with the pixel basis of (a) the object {\bf 3C75}, using  (b) URA, (c) NRA with SD=0.14 and (d) NRA with SD=0.18. The number of sensors is 100.}
\label{fig12}
\end{figure}
}

Some modification of the arguments for Theorems \ref{thm4} and \ref{thm6} leads to  the following coherence bound. 
\begin{theorem}
Assume that the total number of grid point $N$ satisfies 
the bound
\beq
\label{9'}
N\leq {\ep\over 2} e^{K^2/2}
\eeq
with some constants $\delta$ and $K$. Suppose that
the sensor locations $\br_j, j=1,...,n,$ are independent uniform random variables on $[0,1]^2$. 
Then the coherence parameter $\mu$ satisfies the bound
\beq
\mu({  \bA})\leq {|2K^2-1|\over n-1}\label{70}
\eeq
with probability greater than $1-2\ep$.
\label{thm5}
\end{theorem}
In other words, with high probability the coherence parameter for the uniform distribution decays as $n^{-1}$. A central problem in interferometry is the design of an optimal array, see Fannjiang {  2013b} for a discussion from the perspective of compressed sensing.

\commentout{ 
consistent with the numerical result in Fig. \ref{fig11} (a) (red curve with circles). 
A central problem in interferometry is the design of an optimal array.
For comparison, Fig. \ref{fig11} (a) also shows the coherence parameter for the normal distribution with various standard deviations (SD).  The resulting array with normally distributed sensors  is called the
Normal Random Array (NRA). 

As the SD of a NRA  decreases and the array becomes more concentrated, the measurement matrix grows more coherent; conversely, as the array becomes less concentrated, the coherence parameter decreases and approaches that of the URA. The OMP reconstruction shown in Fig. \ref{fig11}(b) largely follows these trends; however, there are two interesting differences.
 
First, the coherence parameter of an NRA does not tend to zero even as the number of sensors grow large. Second, though the coherence parameter curves decrease at a roughly linear rate as the SD of the array increases, the success rate curves do not appear to increase in any clear simple relation to the coherence parameter, 
in contrast to predictions based on  the performance guarantees   \eqref{180}  and \eqref{190}.

Fig. \ref{fig12} shows the OMP reconstruction with 100 sensors of the black hole system {\bf 3C75}. Again, the URA produces an exceptionally accurate reconstruction of the object. The performances of the two NRAs reflect the nuances in sensor density: though an array with SD = 0.14 produces a distinct checkerboard pattern of missed pixels, increasing the SD to just 0.18 erases these artifacts and greatly improves the accuracy of reconstruction. 

For a full account of comparison between URA and NRA as well as between the pixel basis
and the block discrete cosine transform (BDCT), see  Fannjiang 2013b.\\
}

\commentout{
\subsection{Block discrete cosine transform (BDCT)}
 \begin{figure}[t]
\centering 
\includegraphics[width=10cm]{BDCT-mc.pdf}
\caption{Coherence parameter with BDCT versus the number of sensors}
\label{fig15}
\end{figure}
 \begin{figure}[h!]
\centering 
\subfigure[Original object (Crab Nebula)]{
\includegraphics[width=5cm]{nebula1.pdf}
}\hspace{1cm}
\subfigure[URA with pixel basis]{\includegraphics[width=5cm]{nebula-pixel.pdf}
} 
\subfigure[NRA (SD $=0.14$) with BDCT.  
Relative error $=3.78\%$]{
\includegraphics[width=5cm]{bdct-nra.pdf}
}
\hspace{0.8cm}
\subfigure[URA with BDCT. Relative error $=6.33\%$]{
\includegraphics[width=5cm]{nebula-bdct.pdf}
}
\caption{(a) The Crab Nebula and  the OMP reconstruction with (b) URA and pixel basis ,  (c) NRA and BDCT and (d) URA and  BDCT. The number of sensors is 100.}
\label{fig13}
\end{figure}
The block discrete cosine transform (BDCT) is a widely used sparsifying dictionary 
for representing natural images. Following Fannjiang 2013a, we  demonstrate that
 BDCT and random arrays can produce 
 effective sensing matrix for interferometry.
 
The coherence parameter is illustrated in Fig. \ref{fig15}. Clearly,
the use of BDCT  (with $4 \times 4$-pixel transform blocks) significantly reduces the coherence parameter of an 
NRA but does not affect that of the URA.

 Moreover, with BDCT the OMP reconstruction
with the  NRA (SD $=0.14$) in Fig. \ref{fig13}(c) is slightly  accurate than the URA
in Fig. \ref{fig13}(d) even though the former is more coherence than
the latter as shown in Fig. \ref{fig15}. 
On the other hand, the reconstruction with the pixel basis in Fig. \ref{fig13}(b)
is poor since the original object (Crab Nebula) in Fig. \ref{fig13}(a) is not sparse in the pixel basis. 

These observations speak to 1) the importance of coordinating
the  design of 
the sampling scheme with the choice of the sparsifying basis and 2) the fact that 
the coherence parameter does not completely predict the performance of OMP.
}


{\bf Acknowledgements.} Research is supported in part by  US NSF grant DMS-1413373 and Simons Foundation grant 275037.

\bibliographystyle{amsalpha}

\end{document}